\documentclass[a4paper,11pt]{article}
\pdfoutput=1 

\usepackage{jcappub} 

\usepackage{float}
\usepackage{xcolor} 
\usepackage{url}
\usepackage{booktabs,multirow,tabularx, subcaption}

\title{
Impact of Simulation Box Size for Weak Lensing: Replication and Super-Sample Effects
}

\author[a, b, c]{Akira Tokiwa,}
\author[d, e]{Adrian E. Bayer,}
\author[a, b]{Joaquin Armijo,}
\author[a, b]{\\Jia Liu,}
\author[a, b, c]{Ryo Terasawa,}
\author[a, b]{Leander Thiele,}
\author[f, g]{Marcelo Alvarez,}
\author[a, b]{Linda Blot,}
\author[a, b, c]{and Masahiro Takada}

\affiliation[a]{Kavli Institute for the Physics and Mathematics of the Universe (WPI), 5-1-5 Kashiwanoha, Kashiwa-shi, Chiba 277-8583, Japan}
\affiliation[b]{Center for Data-Driven Discovery, Kavli IPMU (WPI), UTIAS, The University of Tokyo, Kashiwa, Chiba 277-8583, Japan}
\affiliation[c]{The University of Tokyo, Department of Physics, 7-3-1 Hongo, Bunkyo-ku, Tokyo 113-0033 Japan}
\affiliation[d]{Center for Computational Astrophysics, Flatiron Institute, 162 5th Avenue, New York NY 10010, USA}
\affiliation[e]{Department of Astrophysical Sciences, Princeton University, Peyton Hall, Princeton NJ 08544, USA}
\affiliation[f]{KIPAC, Stanford University, 452 Lomita Mall, Stanford, CA 94305, USA}
\affiliation[g]{SLAC National Accelerator Laboratory, 2575 Sand Hill Road, Menlo Park, CA 94025, USA}

\emailAdd{akira.tokiwa@gmail.com}
\emailAdd{abayer@princeton.edu}
\emailAdd{joaquin.armijo@ipmu.jp}

\abstract{
We quantify the bias caused by small simulation box size on weak lensing observables and covariances, considering both replication and super-sample effects for a range of higher-order statistics. Using two simulation suites---one comprising large boxes ($3750\,h^{-1}{\rm Mpc}$) and another constructed by tiling small boxes ($625\,h^{-1}{\rm Mpc}$)---we generate full-sky convergence maps and extract $10^\circ \times 10^\circ$ patches via a Fibonacci grid. We consider biases in the mean and covariance of the angular power spectrum, bispectrum (up to $\ell=3000$), PDF, peak/minima counts, and Minkowski functionals. By first identifying lines of sight that are impacted by replications, we find that replication causes a $\mathcal{O}(10\%)$ bias in the PDF and Minkowski functionals, and a $\mathcal{O}(1\%)$ bias in other summary statistics. Replication also causes a $\mathcal{O}(10\%)$ bias in the covariances, increasing with source redshift and $\ell$, reaching $\sim25\%$ for $z_s=2.5$. We additionally show that replication leads to heavy biases (up to $\mathcal{O}(100\%)$ at high redshift) when performing gnomonic projection on a patch that is centered along a direction of replication. We then identify the lines of sight that are minimally affected by replication, and use the corresponding patches to isolate and study super-sample effects, finding that, while the mean values agree to within $1\%$, the variances differ by $\mathcal{O}(10\%)$ for $z_s\leq2.5$. We show that these effects remain in the presence of noise and smoothing scales typical of the DES, KiDS, HSC, LSST, Euclid, and Roman surveys. We also discuss how these effects scale as a function of box size. Our results highlight the importance of large simulation volumes for accurate lensing statistics and covariance estimation.} 

\begin{document}
\maketitle
\flushbottom

\section{Introduction}
Weak gravitational lensing (WL) serves as a powerful probe of dark matter and dark energy. By measuring the subtle distortions in the shapes of distant galaxies, WL thereby captures the gravitational evolution of large-scale structure and offers stringent constraints on key cosmological parameters, such as the amplitude of matter fluctuations ($\sigma_8$) and the matter density parameter ($\Omega_m$), which are often combined as $S_8 \equiv \sigma_8\sqrt{\Omega_m/0.3}$. Recent astronomical surveys, covering both ground-based platforms (e.g., the Dark Energy Survey [DES] \citep{2018ApJS..239...18A} and the Hyper Suprime-Cam Subaru Strategic Program [HSC-SSP] \citep{2018PASJ...70S...4A}) and space-based missions (e.g., {Euclid} \citep{2010arXiv1001.0061R} and the future {Nancy Grace Roman Space Telescope} \citep{2015arXiv150303757S}), are designed to map vast regions of the sky with high depth and precision. These Stage-III and Stage-IV\footnote{The definitions of ``Stage-III'' and ``Stage-IV'' were introduced in the Dark Energy Task Force report \citep{Albrecht2006report}.} initiatives not only test the standard $\Lambda$CDM cosmological model but also address issues such as the Hubble tension and discrepancies in $S_8$ estimates.

To fully leverage the scientific potential of Stage-IV surveys, it is essential to develop robust analysis pipelines that extend beyond conventional two-point statistics. Although the angular power spectrum has traditionally been used to characterize the matter distribution, the inherently non-Gaussian nature of the weak lensing field motivates the incorporation of higher-order statistics, such as the bispectrum, probability density function (PDF), peak/minima counts, Minkowski functionals, and field-level approaches  \citep{2004MNRAS.348..897T, 2014MNRAS.441.2725F, 2015PhRvD..91j3511P, 2018JCAP...10..051F, 2018MNRAS.474..712M, 2020MNRAS.498.4060G, 2021MNRAS.505.2886B, 2022PhRvD.105h3518F, 2023OJAp....6E...1U, 2023arXiv230405928T, 2024MNRAS.528.4513M, Dai:2022dso, Dai:2023lcb, Sharma:2024pth, DES:2024xij, SBI_LSST, NuCNN}. These approaches are more sensitive to the non-Gaussian features of the cosmic web and, when combined with standard two-point measures, offer enhanced constraining power \cite{2023A&A...675A.120E}.

Mock catalogs for WL observables are typically generated by constructing light-cones from cosmological N-body simulations. A common approach involves stacking simulation boxes with periodic boundary conditions side-by-side, or with more complex configurations, thereby reducing computational costs \citep{2010ApJ...709..920S, 2018JCAP...03..049L, 2020JCAP...10..012S, CosmoGridV1, 2024MNRAS.530.5030O}. However, this method introduces systematic artifacts. In particular, the omission of large-scale (super-sample) modes in modestly sized simulation boxes leads to an underestimation of covariance on small scales: this is the super-sample covariance (SSC), a nonlocal contribution to the covariance matrix \citep{2003ApJ...584..702H, PhysRevD.87.123504}. Additionally, replicating simulation boxes to extend the light-cone introduces artificial periodicity, which can bias the mean and variance of the measured summary statistics \citep{2024MNRAS.534.1205C}. Although the impact of SSC on higher-order statistics has been examined in three-dimensional analyses of the matter, halo, and galaxy fields \citep{PhysRevD.108.043521, Schreiner:2024grf}, the effect on higher-order WL observables remains largely unexplored beyond the power spectrum and bispectrum \citep{2018JCAP...06..015B, 2018PhRvD..97d3532C, 2018A&A...611A..83L, 2019A&A...624A..61L, 2023OJAp....6E...1U}, moreover, the combination of super-sample and box replication effects remains unexplored.

To investigate these issues, we compare a light-cone simulated using a big box to a light-cone tiled by many small boxes. Our big ($3750\,h^{-1}{\rm Mpc}$) boxes are taken from the {HalfDome} simulations\footnote{\url{https://halfdomesims.github.io/}}~\citep{2024arXiv240717462B}. We ran small boxes ($625\,h^{-1}{\rm Mpc}$) with identical resolution and other settings to the big box, tiling to achieve the same light-cone volume. By comparing the outputs of these two approaches, we can quantify the impact of super-sample and replication effects on higher-order weak lensing observables. We consider several summary statistics, including the power spectrum, bispectrum, PDFs, peak/minima counts, and Minkowski functionals of the convergence field. This comparative analysis is further refined by incorporating realistic observational conditions: Gaussian shape noise and smoothing are added to the maps at levels appropriate for current and future surveys. Overall, our work aims to disentangle simulation-related errors, thereby providing practical guidelines for the accurate modeling of WL summary statistics and their covariance matrices. This is crucial for reliable cosmological parameter inference, particularly as next-generation surveys (e.g., LSST, {Euclid}, {Roman}) push the boundaries of precision cosmology by probing non-Gaussian features.

This paper is organized as follows. We describe the WL concepts used in this paper in Section \ref{sec:WL}. The setup of the simulations is explained in Section \ref{sec:SimSetup}. We define the set of summary statistics and the calculation of covariance matrices used in this study in Section \ref{sec:Stats}. We show results and discuss in Section \ref{sec:results} before we draw conclusions in Section \ref{sec:conclusions}.

\section{Weak Lensing}\label{sec:WL}
Weak lensing is a powerful cosmological probe of the late-time matter distribution that involves the deflection of light from distant galaxies by the gravitational potential $\Phi$ of intervening matter. Comprehensive reviews of the underlying theory can be found in \citep{2001PhR...340..291B} and \citep{2015RPPh...78h6901K}.  In the thin-lens and small-deflection limit, the mapping between the unlensed source position $\boldsymbol{\beta}$ and the observed image position $\boldsymbol{\theta}$ is described by the Jacobian
\begin{equation}
    \mathbf{A} = \left(
    \begin{array}{cc}
        1 - \kappa - \gamma_1 & -\gamma_2 \\
        -\gamma_2 & 1 - \kappa + \gamma_1
    \end{array}
    \right).
\end{equation}
where $\kappa$ is the convergence, representing the isotropic magnification of images, and $\gamma=\gamma_1+i\gamma_2$ the shear, which accounts for the anisotropic stretching that distorts galaxy shapes \citep{2000ApJ...530..547J, 1993ApJ...404..441K}.
The two are linearly related in Fourier space on the sky by $\tilde\kappa(\boldsymbol{\ell}) = \ell^2\tilde\psi(\boldsymbol{\ell})/2$ and $\tilde\gamma(\boldsymbol{\ell})=\tilde\kappa(\boldsymbol{\ell})e^{2i\varphi_{\ell}}$, where tildes denote Fourier transforms and $\varphi_{\ell}$ is the polar angle of $\boldsymbol{\ell}$ \citep{2000ApJ...530..547J, 1993ApJ...404..441K}.  Although shear is directly observable, we focus on $\kappa$ because it carries identical cosmological information once reconstructed from $\gamma$.

In the weak lensing regime, the deflection angles are sufficiently small that the lensing effect can be described using the Born approximation, wherein light rays are assumed to travel along unperturbed paths. Under this approximation, the convergence $\kappa$ can be written as
\begin{equation}
    \kappa(\hat{\mathbf{n}}) = \int_{0}^{\chi_s}\!\! \mathrm{d}\chi\;
              W(\chi,\chi_s)\,\delta\!\bigl(\hat{\mathbf{n}},\chi\bigr),
    \label{eq:kappa_LOS}
\end{equation}
where the lensing efficiency function $W(\chi,\chi_s)$ is given by
\begin{equation}
 W(\chi,\chi_s)=\frac{3H_{0}^{2}\Omega_{m}}{2c^{2}}\,
                \frac{\chi(\chi_s-\chi)}{a(\chi)\chi_s}.
\end{equation}
 $\delta(\hat{\mathbf{n}}, \chi)$ is the matter density contrast along the line of sight in the direction $\hat{\mathbf{n}}$, and $\chi_s$ is the comoving distance to the source galaxies. 
Equation~\eqref{eq:kappa_LOS} is in good agreement with full multi-plane ray tracing up to $\ell\sim5{,}000$ for the angular power spectrum \citep{2009A&A...499...31H, 2010A&A...523A..28K}. We use the Born approximation to compute convergence maps from our simulations, in line with recent weak lensing analyses of higher-order statistics \cite{Novaes:2024dyh}, leaving studying beyond-Born effects to future work \cite{2017PhRvD..95l3503P}. 

\section{Simulation Setup}\label{sec:SimSetup}

\subsection{N-body Simulations}
\begin{figure}[htbp]
    \centering
    \includegraphics[width=0.8\textwidth]{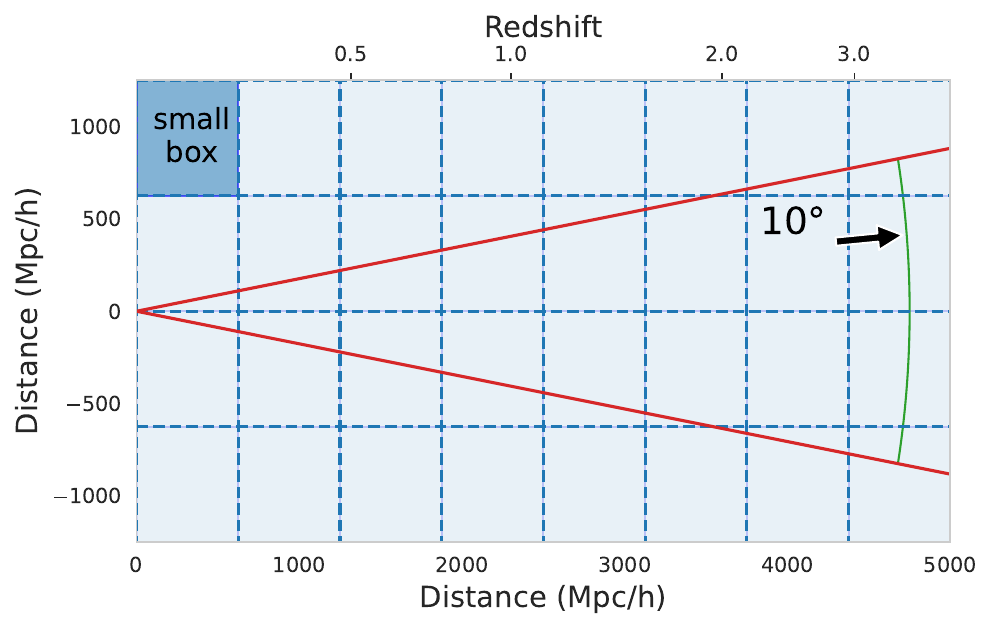}
    \caption[Spatial and redshift setup for the BIGBOX and TILED simulations]{Geometry of the light-cone analysed in this work.  
    The horizontal axis gives the comoving radial distance from the observer (located at the origin), while the corresponding redshift values are shown on the top axis.  
    \textbf{Blue dashed square grid:} the $L_{\mathrm{box}} = 625\,h^{-1}\mathrm{Mpc}$ cubes used in the tiled ensemble.  
    The shaded blue square in the upper left marks one such individual small box.  
    \textbf{Red lines:} the $10^{\circ}$ half-opening angle of the light-cone, which fully fits up to $z\simeq2$ inside a single $L_{\mathrm{box}} = 3750\,h^{-1}\mathrm{Mpc}$ simulation.  
    The green arc illustrates a shell of constant comoving distance, intersecting the $10^{\circ}$ aperture at $z\simeq3$.  
    }
    \label{fig:simulationsetting}
\end{figure}

To investigate the impact of box size on higher-order statistics, we employed two distinct suites of simulations: BIGBOX and TILED.
The BIGBOX simulations are large-volume N-body runs that have no replication effects up to a source redshift $z_s\simeq2$ as the constructed light-cone is contained within the volume of the simulations, and are large enough that the SSC contribution is expected to be sub-percent (see Section. \ref{sec:conclusions} for the $\sigma_b$ scaling and numerical estimate). On the other hand, the TILED simulations cover a volume smaller than the light-cone, thus several small boxes are joined together periodically to match the BIGBOX light-cone volume, introducing (i) replication artifacts, spurious correlations from repeated structures due to periodic tiling, and (ii) super-sample covariance (SSC), i.e., additional covariance from long-wavelength modes larger than the simulated volume.”. We quantify the impact of these effects on higher-order weak lensing observables by comparing the outputs of these two simulation setups. Figure~\ref{fig:simulationsetting} illustrates the spatial and redshift configurations for the BIGBOX and TILED simulations used in this study.

The BIGBOX simulations are taken from the {HalfDome} Project \cite{2024arXiv240717462B}, which comprises a series of large-volume N-body simulations. These simulations were run using the {FastPM} code \citep{10.1093/mnras/stw2123, Bayer_fastpm} with $6144^3$ particles in a box of size $L_{\mathrm{box}}=3750\,h^{-1}\mathrm{Mpc}$ and a particle mass of $m_p = 1.95 \times 10^{10} \, M_{\odot}/h$. The cosmological parameters were adopted from the IllustrisTNG project \citep{2019ComAC...6....2N}. The simulations commenced at an initial redshift of $z = 9$, utilizing an initial linear matter power spectrum at $z = 0$ generated via the \texttt{CLASS} code \citep{2011JCAP...07..034B}. They were evolved over $60$ time steps from $z = 9$ to the present day ($z = 0$). The resulting particle distributions were output in $80$ shells spanning scale factors from $a = 0.2$ to $a = 1.0$, with a uniform spacing of $\Delta a = 0.01$. Within each shell, particles are projected onto a full-sky HEALPix grid \citep{Górski_2005} with $N_{\text{side}} = 8192$, yielding an angular resolution of approximately $0.43$ arcminutes to generate mass maps.

The TILED simulations are constructed by replicating smaller boxes to form the light-cone. These simulations use $1024^3$ particles in a box of size $L_{\mathrm{box}}=625\,h^{-1}\mathrm{Mpc}$, while other settings, such as the particle mass, cosmological parameters, and redshift range remain consistent with those of the BIGBOX simulations. \cite{PhysRevD.108.043521} showed that this box size produces a noticeable super-sample effect in terms of the 3D clustering of matter and halos. Consequently, we expect that such effects will be observable for WL too. Moreover, tiling induces replication effects, 
which introduce artificial correlations and anisotropies in the mass maps, biasing measurements.

In total, $11$ realizations of the BIGBOX simulation and $20$ realizations of the TILED simulation are used, each initialized with a distinct random seed. Note that the observer is positioned at a cube corner shared by eight replicas, which aligns some lines of sight with box faces/edges; these sightlines are most susceptible to replication artifacts.
Thus, in our analysis, we first assess the replication effects and super-sample effects together, determining the particular directions in the sky that are most heavily affected by replication effects, and then exclude these regions from our analysis to isolate and study super-sample effects.

\subsection{Convergence-Map Patch Extraction}

To construct the convergence maps from the projected mass maps, we employ the Born approximation to calculate the convergence field $\kappa$ at each redshift. We considered source redshifts of $z_s = [0.5, 1.0, 1.5, 2.0, 2.5]$, covering distances relevant for both current Stage-III and upcoming Stage-IV surveys..

To mimic realistic observational conditions, we add Gaussian shape noise to the convergence maps. The shape noise is characterized by
\begin{equation}
    \sigma_{\kappa, \text{noise}}^2 = \frac{\sigma_{\epsilon}^2}{2\, n_{\mathrm{gal}}\, A_{\mathrm{pix}}},
\end{equation}
where $\sigma_{\epsilon}$ is the intrinsic ellipticity dispersion of galaxies, set to $\sigma_{\epsilon} = 0.26$ \citep{2019A&A...627A..59E}, $n_{\mathrm{gal}}$ is the galaxy number density per square arcminute, and $A_{\mathrm{pix}}$ is the solid angle of a pixel, set to $(0.29\,\text{arcmin})^2$. We considered four different survey setups (DES/KiDS, HSC, LSST/Euclid, and Roman) with galaxy number densities of $n_{\mathrm{gal}} = 7,\,15,\,30,\,50$ galaxies per square arcminute, respectively. Table~\ref{tab:survey_comparison} provides a comprehensive overview of these six pivotal surveys focusing on their observational capabilities. 
\begin{table}[t]
    \centering
    \begin{tabular}{lccc}
         \hline
        \textbf{Survey}  & \textbf{Galaxy Density (arcmin$^{-2}$)} & \textbf{Median Redshift} \\ \hline
        DES/KiDS  &  7 & 0.4 \\
        HSC &  15 & 0.7 \\
        LSST/{Euclid} & 30 & 1.0 \\
        {Roman}  & $50$ & 1.5 \\
         \hline
    \end{tabular}
     \caption{Assumed characteristics of ongoing and upcoming weak lensing surveys studied in this work. Note that we assume the number density for the case of no tomography---splitting into tomographic bins would increase the noise.
     }
    \label{tab:survey_comparison}
\end{table}
Unless stated otherwise, quoted galaxy number densities $n_{\rm gal}$ correspond to the no-tomography case
(a single, broad source bin). If the samples are partitioned into $N_{\rm bin}$ tomographic bins,
the per-bin effective number density scales as $n_{\rm gal}\!\rightarrow n_{\rm gal}/N_{\rm bin}$ and the shape-noise level increases accordingly.

\begin{figure}[t]
    \centering
    \includegraphics[width=0.32\textwidth]{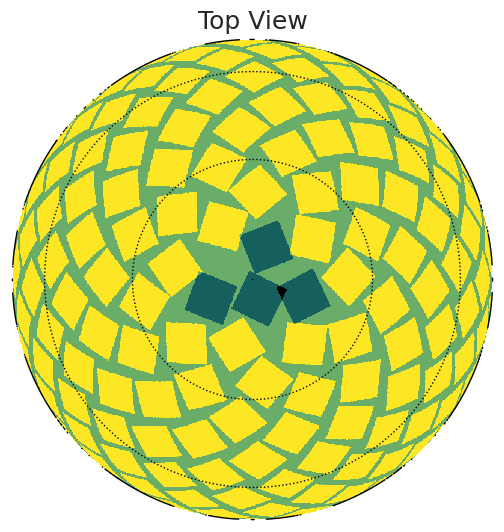}
    \includegraphics[width=0.32\textwidth]{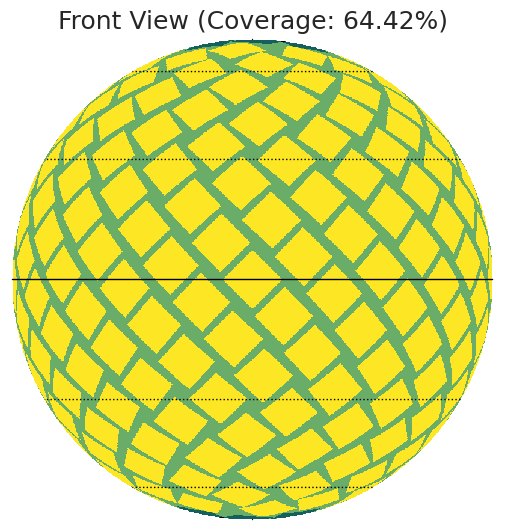}
    \includegraphics[width=0.32\textwidth]{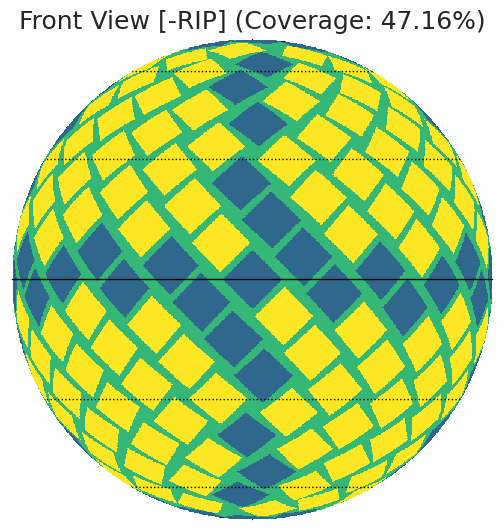}
    \caption[Visualization of the Fibonacci grid]{Visualization of the Fibonacci grid with $N_{\text{patches}} = 273$, where each patch covers approximately $10\times 10\, {\rm deg}^2$. After optimization and masking, the number of patches is reduced to $N_{\text{patches}} = 194$, effectively covering $47\%$ of the sky. Each panel shows the patch distribution from a top view (left), front view (middle), and the front view after masking replication-influenced patches (RIPs) (right). The black patches in the left and middle panels correspond to tiles that are excluded from all analyses, as the HEALPix grids near poles are highly skewed and cause unstable results after gnomonic projection to the 2D plane; the blue tiles on the right panel correspond to the additional tiles that are excluded from parts of our analysis due to being replication-impacted. The patch in the center of the right panel (at $(\theta=\pi/2, \phi=0)$) is especially affected by replication effects.} 
    \label{fig:fibonacci}
\end{figure}

To simplify our analysis by working on a flat projection, we extracted $10\times 10\, {\rm deg}^2$ patches from the full-sky convergence maps. These patches were obtained by projecting the data from the HEALPix grid onto a Euclidean grid using the \texttt{healpy.gnomview(..., return\_projected\_map=True)} function in Python \cite{2019JOSS....4.1298Z}. This is similar to the setup of \cite{Euclid:2019zmk} and has been shown to work well for two-point statistics in \cite{Pires:2008dw}. To maximize the number of uniformly distributed, non-overlapping patches, we employed a Fibonacci grid \citep{2006QJRMS.132.1769S, 2023MNRAS.524.5591F} for patch extraction. The center of each patch was set at the vertices of the Fibonacci grid, defined by golden ratio spirals:
\begin{equation}
    \sin \theta_i = \frac{2i}{2N + 1}, \quad \phi_i = \frac{2 \pi i}{\varphi}, \quad -N \leq i \leq N, \quad 0 \leq \theta_i \leq \pi,
\end{equation}
where $N$ denotes the number of patches and $\varphi = (1+\sqrt{5})/2$ is the golden ratio. Figure~\ref{fig:fibonacci} illustrates the Fibonacci grid.

Following \cite{2023MNRAS.524.5591F}, the maximum number of patches is obtained by aligning the diagonals of square patches with the longitude lines. Initially, optimization and masking yielded $N_{\text{patches}} = 273$. We remove the 8 patches near the poles to avoid projection artifacts associated with poles, 
reducing the number to $N_{\text{patches}} = 265$  patches, each $10\times 10\, {\rm deg}^2$ in size, corresponding to a sky coverage of $64\%$. After analyzing the results in the presence of large replication effects (in Section \ref{sec:boxreplication}), we then exclude patches containing regions with heavy replication along the line of sight (see right panel of Figure~\ref{fig:fibonacci}) to better isolate and study the super-sample effect (in Section \ref{sec:ssc}). This results in a sample of $N_{\text{patches}} = 194$ for studying super-sample effects, effectively covering $47\%$ of the sky, a coverage sufficiently large for the Euclid and LSST surveys and comparable to that reported in \cite{2019MNRAS.488.5833D}.
Each patch is represented by a $2048 \times 2048$ pixel grid, corresponding to a pixel size of $0.29$ arcmin. In total, this procedure yields 2134 patches from the BIGBOX simulations and 3880 patches from the TILED simulations (194 patches per realization). For each realization, the covariance was computed using these 194 patches extracted from the full-sky map.

To mitigate noise and pixelization effects, we apply a Gaussian smoothing filter to the convergence maps. The Gaussian filter is defined as:
\begin{equation}
    G(\theta) = \frac{1}{\pi \theta_{\mathrm{G}}^2}
            \exp\!\Bigl(-\frac{\theta^{2}}{2\theta_{\mathrm{G}}^{2}}\Bigr),
\end{equation}
where $\theta$ denotes the angular distance from the center in units of arcminutes of the filter and $\theta_{\mathrm{G}}$ is the smoothing scale. For our analysis, we consider $\theta_{\mathrm{G}}$=2, 5, 8, 10 arcmin---using 2 arcmin unless otherwise state. By convolving the noisy convergence map with the Gaussian filter, we obtained the smoothed convergence map:
\begin{equation}
    \kappa_{\mathrm{smoothed}}(\hat{\mathbf{n}}) = \int  \mathrm{d}\Omega' \, G\big(|\hat{\mathbf{n}} - \hat{\mathbf{n}}'|\big) \kappa_{\mathrm{obs}}(\hat{\mathbf{n}}').
\end{equation}
The smoothing kernel effectively suppresses small-scale structures, thereby altering the range of $\kappa$ values. 
We normalize the $\kappa$ values by the standard deviation $\sigma_{\kappa}$ of each patch's convergence map.

\section{Statistics}\label{sec:Stats}
In this section, we introduce the statistical measures used to quantify the weak lensing signal. These measures include the angular power spectrum, bispectrum, peak and minima counts, probability density function (PDF), and Minkowski functionals. Collectively, these statistics provide comprehensive insights into both the Gaussian and non-Gaussian features of the large-scale structure.

\subsection{Summary Statistics}

\subsubsection{Angular Power Spectrum}
The angular power spectrum, $C_{\ell}$, 
is computed from the un-smoothed convergence maps using \texttt{lenstools}~\cite{2016A&C....17...73P}. We consider a multipole range from $\ell = 300$ to $\ell = 3000$ in eight logarithmically-spaced bins. This binning scheme is chosen to maintain consistency with the multipole selection used in the Hyper Suprime-Cam Year 3 (HSC Y3) cosmic shear analysis \cite{2023PhRvD.108l3519D}. The lower limit of $\ell = 300$ excludes large-scale modes, as our primary focus is on higher-order statistics that are most sensitive to intermediate and small angular scales. Conversely, the upper limit of $\ell = 3000$ ensures the inclusion of small-scale modes, where the super-sample effect becomes increasingly significant. 
We validated our measurements against the \texttt{Halofit} model \cite{2012ApJ...761..152T} and found an overall good agreement. 

\subsubsection{Bispectrum}
The bispectrum, $B_{\ell}$, is computed from the un-smoothed convergence maps using \texttt{lenstools}. We consider three distinct configurations: equilateral ($\ell_1 = \ell_2 = \ell_3$), squeezed ($\ell_1 = \ell_2 = 10\ell_3$), and isosceles ($\ell_1 = \ell_2 = 2\ell_3$). These configurations are chosen to capture different bispectrum shapes and provide complementary information about the underlying matter distribution. The bispectrum is computed over the same multipole range as the angular power spectrum, namely $\ell \in [300, 3000]$ in eight logarithmically-spaced bins. 

\subsubsection{Probability Density Function (PDF)}
We compute the probability density function (PDF) using \texttt{lenstools} on the smoothed convergence maps. Each PDF is derived over a normalized range of $-4 \leq \nu\equiv \kappa/\sigma_{\kappa} \leq 4$ in eight linearly-spaced bins, where $\sigma_{\kappa}$ is the standard deviation of each individual patch. This range is chosen to align with the study by \cite{2023arXiv230405928T}, and the binning is optimized to ensure that each bin contains a sufficient number of data points. 


\subsubsection{Peak and Minima Counts}
Peak and minima counts are computed from the smoothed convergence maps using \texttt{lenstools}. The counts are derived over the same normalized range as the PDF, $-4 \leq \nu\equiv\kappa/\sigma_{\kappa} \leq 4$, divided into eight linearly-spaced bins. Although the extreme bins (corresponding to the highest peaks and lowest minima) contain relatively few data points, they are retained to ensure consistency with the PDF binning scheme.

\subsubsection{Minkowski Functionals}
The Minkowski functionals are calculated from the smoothed convergence maps using a custom implementation following \cite{2012PhRvD..85j3513K}. They are computed as:
\begin{align}
    V_0(\nu_0) &\approx \frac{1}{N_{\mathrm{pix}}} \sum_{i=1}^{N_{\mathrm{pix}}} \Theta(\nu_i - \nu_0), \label{eq:V0_discrete} \\
    V_1(\nu_0) &\approx \frac{1}{4N_{\mathrm{pix}}} \sum_{i=1}^{N_{\mathrm{pix}}} \sum_{j \in \mathcal{N}(i)} \left|\Theta(\nu_i - \nu_0) - \Theta(\nu_j - \nu_0)\right|, \label{eq:V1_discrete} \\
    V_2(\nu_0) &\approx \frac{1}{2\pi N_{\mathrm{pix}}} \sum_{i=1}^{N_{\mathrm{pix}}} \delta_D(\nu_i - \nu_0) \left(\frac{\nu_{,xx} \nu_{,yy} - \nu_{,xy}^2}{\nu_{,x}^2 + \nu_{,y}^2}\right), \label{eq:V2_discrete}
\end{align}
where $\Theta$ is the Heaviside step function, $\delta_D$ is the Dirac-delta function, $\mathcal{N}(i)$ denotes the neighboring pixels of pixel $i$, and the derivatives are estimated via finite differences. As with the PDF and peak/minima counts, these functionals are computed over the normalized range $-4 \leq \nu\equiv\kappa/\sigma_{\kappa} \leq 4$, divided into eight linearly spaced bins.

\subsection{Covariance Matrix}
The covariance matrix encapsulates the uncertainties and correlations among different measurements. It plays a critical role in the parameter estimation techniques, including maximum likelihood analyses, Bayesian inference, and forecasting the capabilities of future surveys via the Fisher information matrix. The covariance between two observables, $\mathcal{O}_i$ and $\mathcal{O}_j$, is defined as:
\begin{equation}
    \mathrm{Cov}(\mathcal{O}_i, \mathcal{O}_j) = \left\langle \left(\mathcal{O}_i - \langle \mathcal{O}_i \rangle\right)\left(\mathcal{O}_j - \langle \mathcal{O}_j \rangle\right) \right\rangle,
\end{equation}
where $\langle \cdot \rangle$ denotes the ensemble average over multiple realizations. For an unbiased estimator under the assumption of independent samples, the covariance matrix is computed as:
\begin{equation}
    \label{eq:covariance}
    \mathrm{Cov}(\mathcal{O}_i, \mathcal{O}_j) = \frac{1}{N_{\mathrm{sim}} - 1} \sum_{n=1}^{N_{\mathrm{sim}}} \left(\mathcal{O}_i^{(n)} - \langle \mathcal{O}_i \rangle\right) \left(\mathcal{O}_j^{(n)} - \langle \mathcal{O}_j \rangle\right),
\end{equation}
where $N_{\mathrm{sim}}$ is the number of simulations, and $\mathcal{O}_i^{(n)}$ denotes the $i$-th observable in the $n$-th simulation.

Capturing the physical correlations and avoiding numerical correlations is crucial when estimating the covariance, as any bias in the covariance could cause biases, or incorrect confidence intervals, in cosmological parameters inference. Box size effects can both underestimate the correlations, as too small of a box will ignore important correlations between large-scale modes and small-scales modes (this is the super-sample effect), and overestimate the correlations as tiling introduces numerical correlations due to repeated structures appearing when boxes are replicated. This affects both the diagonal (variance) and the off-diagonal (covariance) terms \cite{2009MNRAS.396...19N, PhysRevD.108.043521}. 
%
While the super-sample effect has been studied for the angular power spectrum and bispectrum in the past~\citep{2018JCAP...06..015B, 2018PhRvD..97d3532C, 2018A&A...611A..83L, 2019A&A...624A..61L, 2023OJAp....6E...1U}, no study, until now, has been performed for the other higher-order statistics used in cosmology as these require large-scale cosmological simulations.

We compute the correlation matrix for each summary statistic to analyze the interdependence among different scales and configurations. The correlation matrix is defined as:
\begin{equation}
    \rho_{ij} = \frac{\mathrm{Cov}(\mathcal{O}_i, \mathcal{O}_j)}{\sqrt{\mathrm{Cov}(\mathcal{O}_i, \mathcal{O}_i)\,\mathrm{Cov}(\mathcal{O}_j, \mathcal{O}_j)}},
\end{equation}
where $\mathcal{O}_i$ and $\mathcal{O}_j$ represent the $i$-th and $j$-th bin of a summary statistic, or of a collection of summary statistics. The correlation matrix normalizes off-diagonal elements by the scale-dependent variance, thereby facilitating direct comparisons between different simulations and summary statistics.

To assess the impact of the replication and super-sample effects, we compare the covariance matrices derived from the BIGBOX and TILED simulations by calculating their ratios:
\begin{equation}
    R^{\mathrm{Cov}}_{ij} = \frac{\mathrm{Cov}^{\mathrm{BIGBOX}}_{ij}}{\mathrm{Cov}^{\mathrm{TILED}}_{ij}}, \quad R^{\rho}_{ij} = \frac{\rho^{\mathrm{BIGBOX}}_{ij}}{\rho^{\mathrm{TILED}}_{ij}},
\end{equation}
where $\mathrm{Cov}^{\mathrm{BIGBOX}}$ and $\mathrm{Cov}^{\mathrm{TILED}}$ denote the covariance matrices from the BIGBOX and TILED simulations, respectively, and $\rho^{\mathrm{BIGBOX}}$ and $\rho^{\mathrm{TILED}}$ represent the corresponding correlation matrices. 
The choice of BIGBOX on the numerator is according to the common convention in SSC analyses.

\section{Results} \label{sec:results}

\begin{figure}
    \centering
    \includegraphics[width=0.98\textwidth]{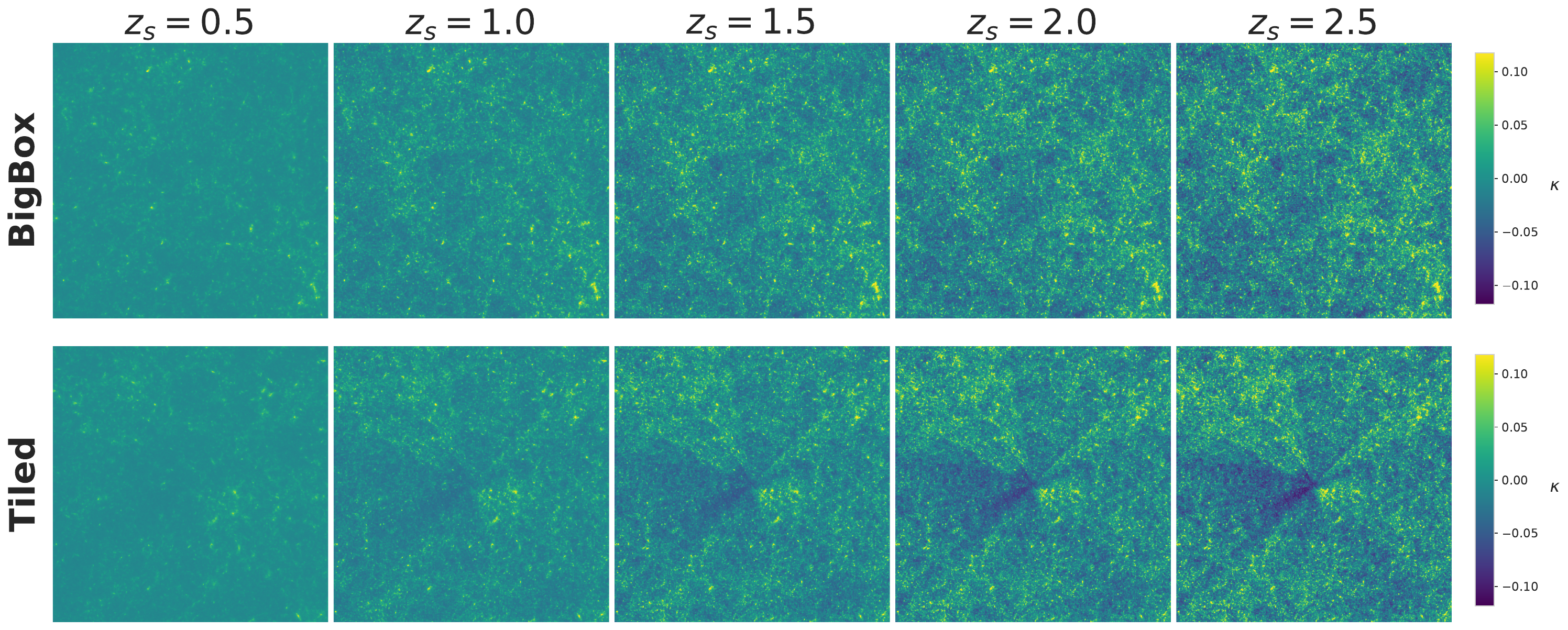}
    \caption[Replication kaleidoscope pattern in tiled box]
    {
        Illustration of convergence maps at different redshifts (columns) for the BIGBOX (top) and TILED box (bottom). A clear kaleidoscope pattern can be observed in the TILED box, which becomes larger at higher source redshift, due to replication effects. Such correlated structures do not appear in the BIGBOX.
    }
    \label{fig:rep_illustration}
\end{figure}

\begin{figure}
    \centering
    \includegraphics[width=0.98\textwidth]{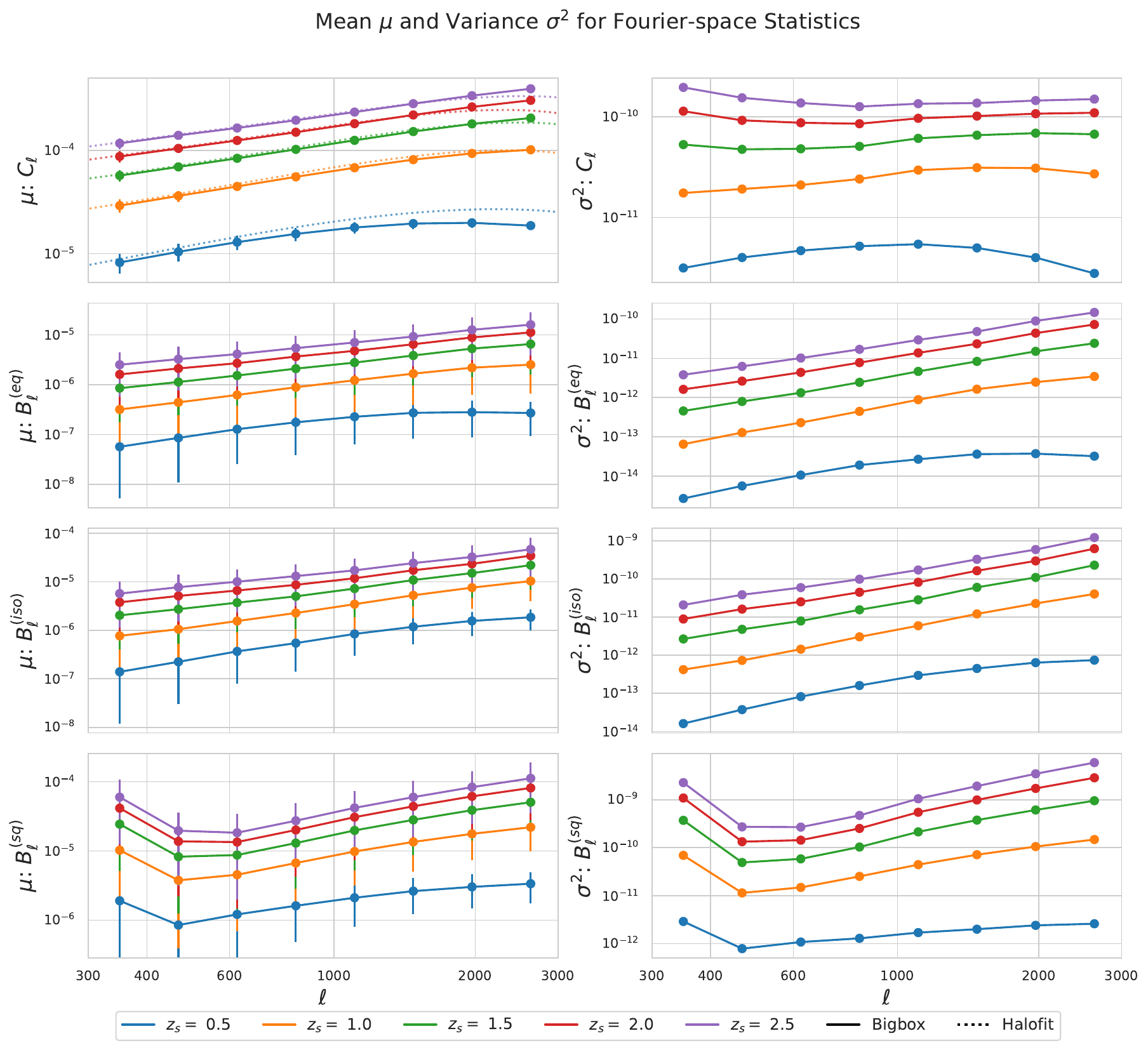}
    \caption[Mean and Variance for $C^{\kappa\kappa}_{\ell}$ and Bispectra in BIGBOX]
    {
        Mean (left panels) and variance (right panels) in the BIGBOX simulation of the angular power spectrum ($C_{\ell}$) and the bispectrum components across three configurations: equilateral ($B_{\ell}^{\text{(eq)}}$), isosceles ($B_{\ell}^{\text{(iso)}}$), and squeezed ($B_{\ell}^{\text{(sq)}}$) as function of multipole $\ell$. Results are presented for source redshifts $z_s = 0.5$ (blue), $1.0$ (orange), $1.5$ (green), $2.0$ (red), and $2.5$ (purple). A comparison with $C_\ell$ from Halofit is given by the dotted line. Note, the power spectrum and bispectrum results are normalized by factors of $\ell(\ell+1)/2\pi$ and $\ell^4$ respectively. 
    }
    \label{fig:mean_std_ell}
\end{figure}

\begin{figure}
    \centering
    \includegraphics[width=0.98\textwidth]{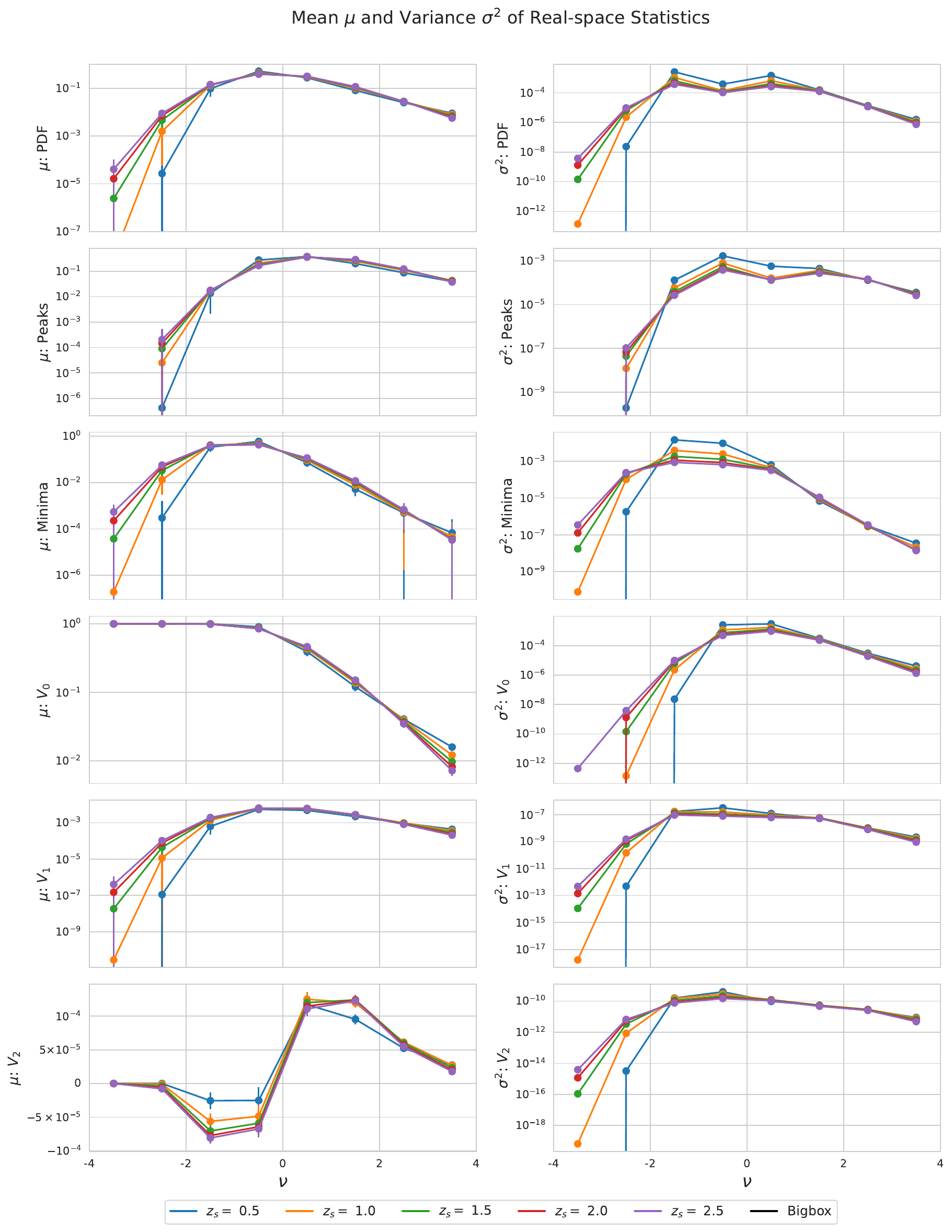}
    \caption[Mean and Variance for PDF, Peak/Minima Counts, and Minkowski Functionals in BIGBOX]
    {
        Same as Figure~\ref{fig:mean_std_ell} but for $\nu$-binned statistics, including the the probability density function (PDF), peak and minima counts, and Minkowski Functionals ($V_0$, $V_1$, $V_2$). 
    }
    \label{fig:mean_std_nu}
\end{figure}

We now present our results comparing the mean and (co)variance of all summary statistics in the BIGBOX and TILED simulation suites. Unless otherwise specified, the results presented are based on noiseless simulations with a smoothing scale of $2\,\mathrm{arcmin}$. The power spectrum and bispectrum results are normalized by factors of $\ell(\ell+1)/2\pi$ and $\ell^4$, respectively, to facilitate clear comparisons. 

We start by considering box size effects in the presence of replication and super-sample effects in Subsection \ref{sec:boxreplication}, where we quantify the magnitude of the replication effects. We then remove patches of the sky that are heavily affected by replication to study the super-sample effect in Subsection \ref{sec:ssc}.
Finally, we show results when including pixel noise and smoothing filters to mimic data for current and future survey configurations in Subsection \ref{sec:survey}.

Figure \ref{fig:rep_illustration} shows convergence maps for the BIGBOX and TILED simulations at various source redshifts. A clear kaleidoscope pattern is observed in the TILED box, which becomes larger at higher source redshift, due to replication effects. We now proceed to quantify the effect this has on the mean and variances of different summary statistics.

Figures \ref{fig:mean_std_ell} and \ref{fig:mean_std_nu} show the mean and variance of each summary statistic in the BIGBOX for both Fourier space and real space statistics respectively. Throughout this section we will study how the TILED means and variances compare by considering the ratio for each statistic between the BIGBOX and TILED simulations. We also analyze the off-diagonal components of the covariance by presenting the covariance- and correlation-matrix ratios for each summary statistic in Appendix \ref{app:cor}. 

\subsection{Replication Effect} \label{sec:boxreplication}

\begin{figure}[t]
    \centering
    \includegraphics[width=\textwidth]{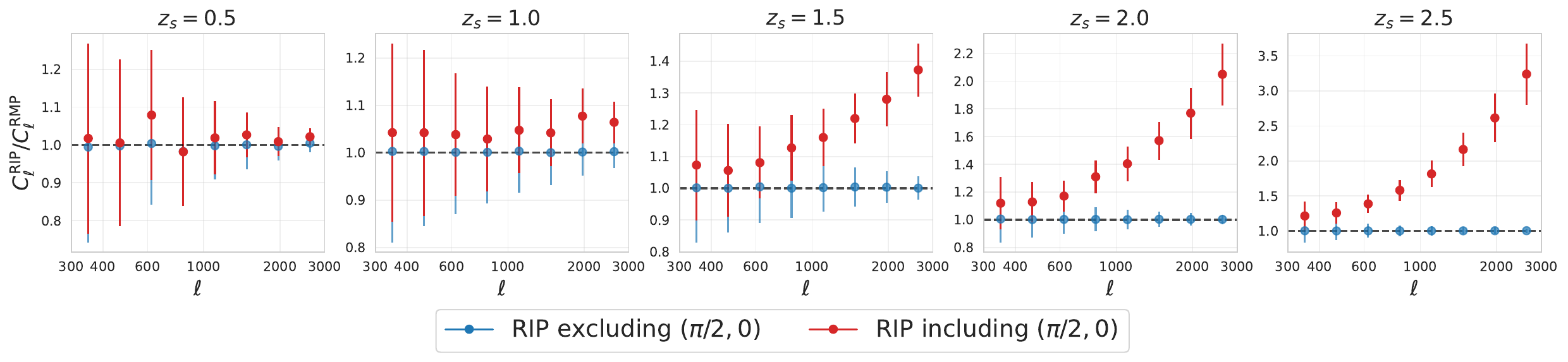}
    \caption[Angular Power Spectrum for Heavily Influenced RIPs]
    {\label{fig:boxreplication_heavy}
        Angular power spectrum ratio for RIPs versus RMPs from the TILED simulations. We consider RIPs with the $(\pi/2,0)$ point (red) and without it (blue). While most RIPs differ only marginally from RMPs, patches centered on $(\pi/2,0)$ exhibit a pronounced difference in both mean and variance due to the gnomonic projection.}
\end{figure}

\begin{figure}[t]
    \centering 
    \includegraphics[width=\textwidth]{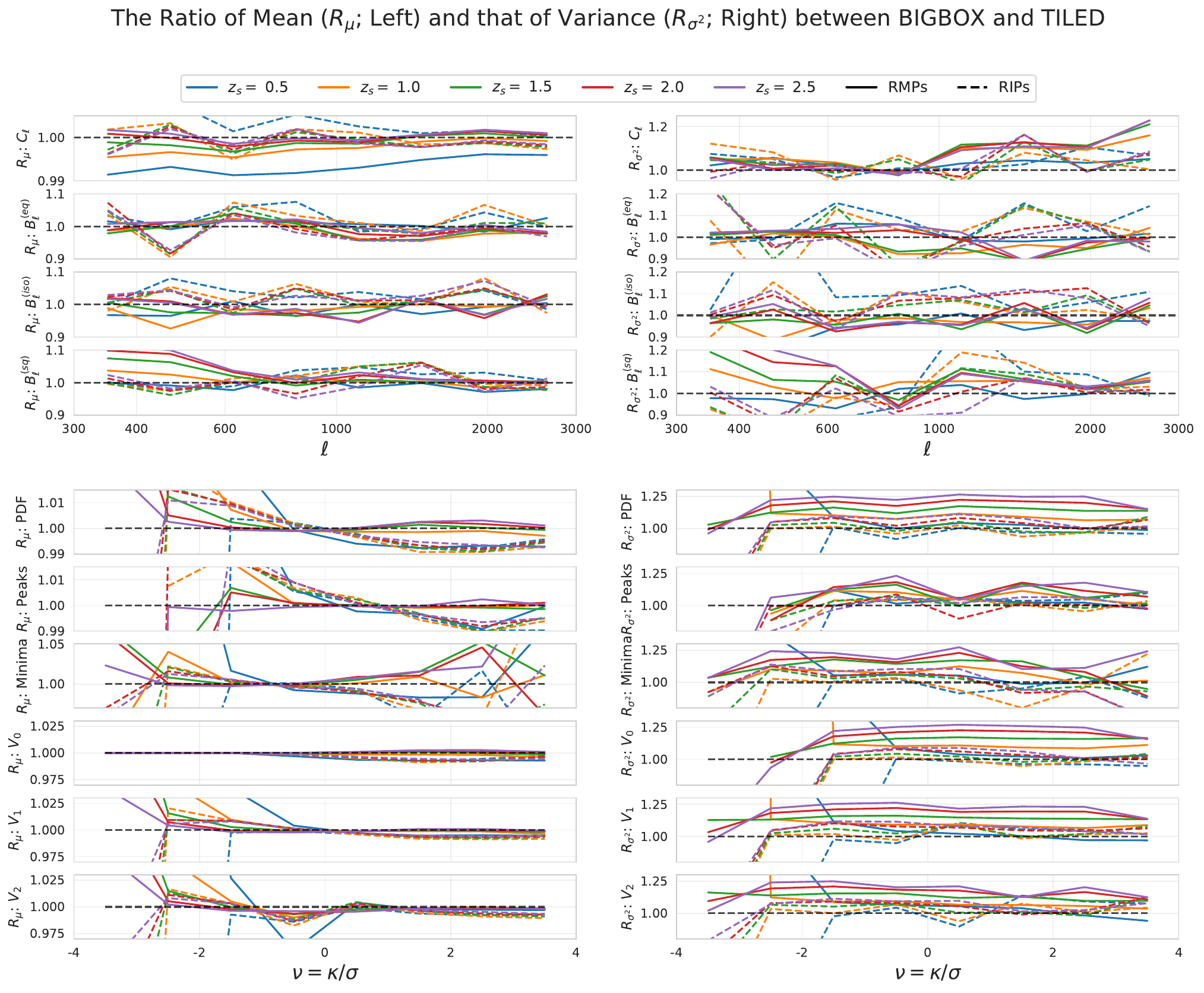} 
    \caption[BIGBOX/TILED Ratios of Mean and Variance for RIPs and RMPs]
    {
        Ratios of the mean (left panels) and variance (right panels) 
        of summary statistics between the BIGBOX and TILED simulations for both Replication-Influenced Patches (RIP; dashed) and Replication-Minimal Patches (RMP; solid).
    }
    \label{fig:boxreplication_main}
\end{figure}

To systematically investigate the replication effect, we categorize simulation regions into two distinct groups: regions significantly affected by artifacts, termed \textbf{Replication-Influenced Patches (RIP)}, and regions minimally affected by artifacts, termed \textbf{Replication-Minimal Patches (RMP)}. The criteria for categorizing patches are as follows:
\begin{itemize}
    \item \textbf{Replication-Influenced Patches (RIP)}:
    \begin{itemize}
        \item {Patches near the equator}: 
        \[
            \left| \theta_i - \frac{\pi}{2}\right| \leq R_{\text{patch}}
        \]
        \item {Patches near the edges of octants}: 
        \[
            \left| \phi_i - \frac{k\pi}{2} \right| \leq R_{\text{patch}} \quad \text{for} \quad k = 0, 1, 2, 3
        \]
    \end{itemize}
    \item \textbf{Replication-Minimal Patches (RMP)}:
    \begin{itemize}
        \item All other patches that are not RIP.
    \end{itemize}
\end{itemize}
Here, \((\theta_i, \phi_i)\) denotes the center of patch \(i\), and \(R_{\text{patch}} = 5\sqrt{2}\,\mathrm{deg}\) is the half-diagonal of the patch. RIPs are shown in blue in the right panel of Figure \ref{fig:fibonacci} and are determined by the fact that they lie along the axes of symmetry of the tiled box, implying pronounced repeated structures.

Figure \ref{fig:boxreplication_heavy} shows the angular power spectrum ratio between RIPs and RMPs from the TILED simulations. We separately consider RIPs with the $(\pi/2,0)$ point (red) and without it (blue). While most RIPs differ only marginally from RMPs, patches centered on $(\pi/2,0)$ exhibit a significant deviation in both the mean and variance of summary statistics compared to other RIPs. 
Since we are using gnomonic projection \cite{2002A&A...395.1077C} to project the patches, values near the center of the patch are affected by the replication effect most; from the definition of the Fibonacci  ratio,  $(\pi/2,0)$ is at the center of a patch, while other points (such as $(\pi/2, \pi/2)$) are not at the center of a patch and thus do not have a large bias. It would be interesting further work to investigate the impact of this effect on other statistics, but for the purpose of this work, we will drop these patches for the remainder of the analysis.

Thus, to analyze the difference between RIPs and RMPs more clearly, we remove the RIP centered on $(\pi/2,0)$. After these adjustments, we obtained 70 RMPs per realization, resulting in a total of 1,400 RMPs for the 20 TILED simulations and 770 RMPs for the 11 BIGBOX simulations across all realizations.

Figure~\ref{fig:boxreplication_main} presents the ratios of the mean and variance between BIGBOX and TILED for various summary statistics, separating the RIPs and RMPs.
The analysis reveals that the mean angular power spectra in RIPs are $\sim2\%$ higher than those in RMPs due to higher correlations in RIPs. 
While we focus on $\nu\equiv\kappa/\sigma$ as the independent variable for real-space statistics in the main text, we consider different combinations of mean subtraction and variance scaling in Appendix \ref{app:nu_kap}.

For RIPs, the BIGBOX $C_\ell$ is larger than the TILED $C_\ell$, due to cosmic variance---in Appendix \ref{app:nu_kap} we see that after subtracting the mean $\kappa$ from the maps this effect is removed.
The mean value of $\nu$-binned statistics in RIPs from the TILED simulation are higher in positive \(\nu\) bins and lower in negative \(\nu\) bins compared to the BIGBOX. 
This is because the PDF is skewed in the $\nu>0$ direction in the absence of repetition (see Figure \ref{fig:mean_std_nu}), thus replication will cause an increase in the occurrence of positive $\nu$ values and decrease in negative $\nu$ values.
It is also notable that the bias due to replication (i.e.~the difference between RIPs and RMPs) is largest for lower $z_s$: one might have thought that the replication effect should grow with $z_s$ as there is more replication along the line of sight, however, the size of the patch in the direction perpendicular to the line of sight (or, the cross section area) also increases with $z_s$, washing out the impact of the repeated structure.

In terms of the variance, there is a notable trend for $\nu$-binned statistics of RIP patches being similar between BIGBOX and TILED. This is because of the division by $\sigma$ in $\nu=\kappa/\sigma$ somewhat accounting for the variance introduced by replication in the summary statistic. Appendix \ref{app:nu_kap} shows that when computing in bins of $\kappa$ there is a considerable variance ratio in RIPs. Similarly subtracting the mean $\kappa$ value increases the difference in variance in RIPs.

The entire correlation and covariance matrices for RIPs are discussed in Appendix \ref{app:cor_rep}.

\subsection{Super-Sample Effect} \label{sec:ssc}

Having studied the covariance in RIPs, we now focus on the RMPs of Figure \ref{fig:boxreplication_main} to isolate the super-sample effect.

For most summary statistics, deviations in the ratio of the mean between the TILED box and BIGBOX are less than 1\%, underscoring the robustness of both simulation pipelines in reproducing the mean of the statistics. The $\sim1\%$ difference in $C_\ell$ is due to window effects associated with power computation \cite{2014PhRvD..89h3519L}. 
However, systematic deviations between the TILED and BIGBOX simulations emerge at low $\nu$ values for peak/minima counts and Minkowski functionals, and at high $\nu$ for minima. These deviations likely result from the limited box size of the TILED simulations in capturing extremely high-density and low-density regions, leading to fewer data points in the extreme tails of the distribution. Appendix \ref{app:nu_kap} shows that the ratio in the mean for real-space statistics between the TILED and BIGBOX simulations is much larger when binning in $\kappa$ instead of $\nu$.

The variance  of the angular power spectrum is significantly larger in the BIGBOX at high $\ell$. This increase is consistent with the expected amplification due to the inclusion of large-scale (super-sample) modes in the BIGBOX simulations, which are absent in the TILED simulations. Furthermore, discrepancies in the variance ratios grow with increasing source redshift, reflecting the progressive loss of large-scale modes in the TILED simulations. For $\nu$-binned statistics, the BIGBOX simulations consistently yield higher variances across nearly all $\nu$ bins, with this trend becoming more pronounced at higher source redshifts. These findings indicate a $>10\%$ bias in the covariance, underscoring the necessity of accounting for the super-sample effect to ensure accurate cosmological parameter estimation from weak lensing statistics.
The bispectrum is the only notable exception, which has as small super-sample effect, as shown in \cite{PhysRevD.108.043521}, except in the squeezed case where the squeezed side of the triangle introduces a super-sample effect.
Appendix \ref{app:nu_kap} shows that these $\mathcal{O}(10\%)$ effects remain no matter which binning one chooses.

The entire correlation and covariance matrices for RMPs are discussed in Appendix \ref{app:cor_ssc}.

\subsection{Effects of Noise and Smoothing Scales} \label{sec:survey}

\begin{figure}[t]
    \centering 
    \begin{subfigure}{0.5\textwidth}
        \centering
        \includegraphics[width=\textwidth]{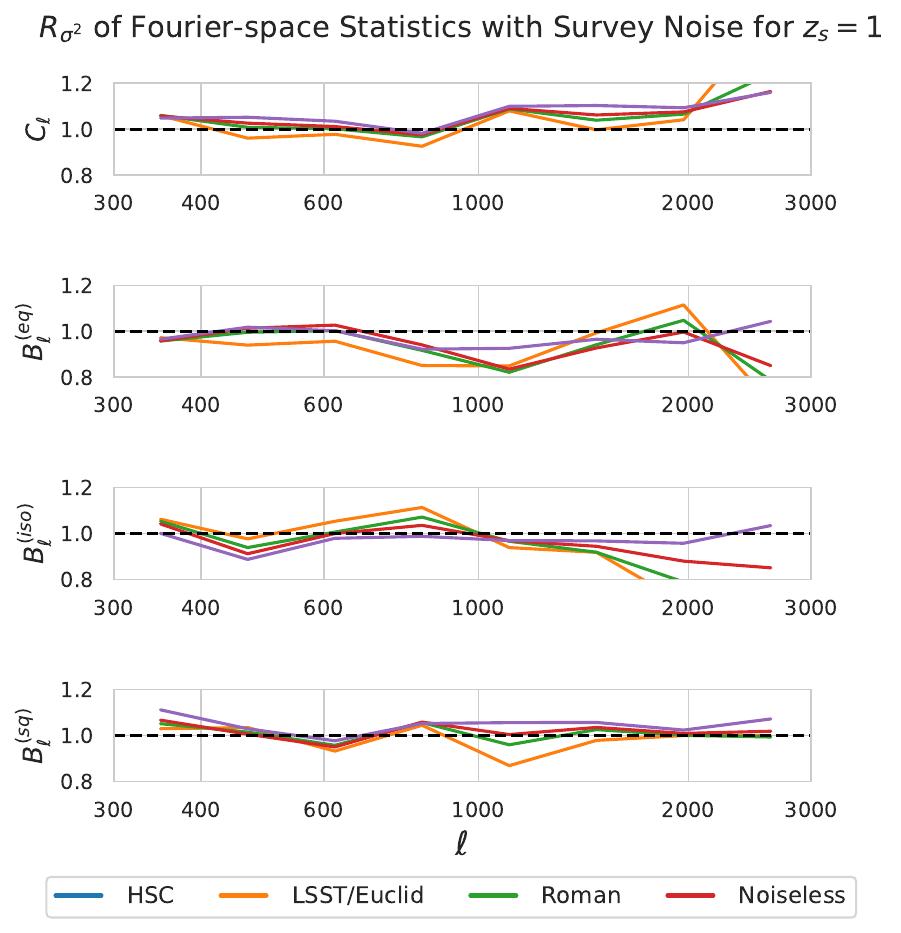} 
    \end{subfigure}
    \hfill
    \begin{subfigure}{0.46\textwidth}
        \centering
        \includegraphics[width=\textwidth]{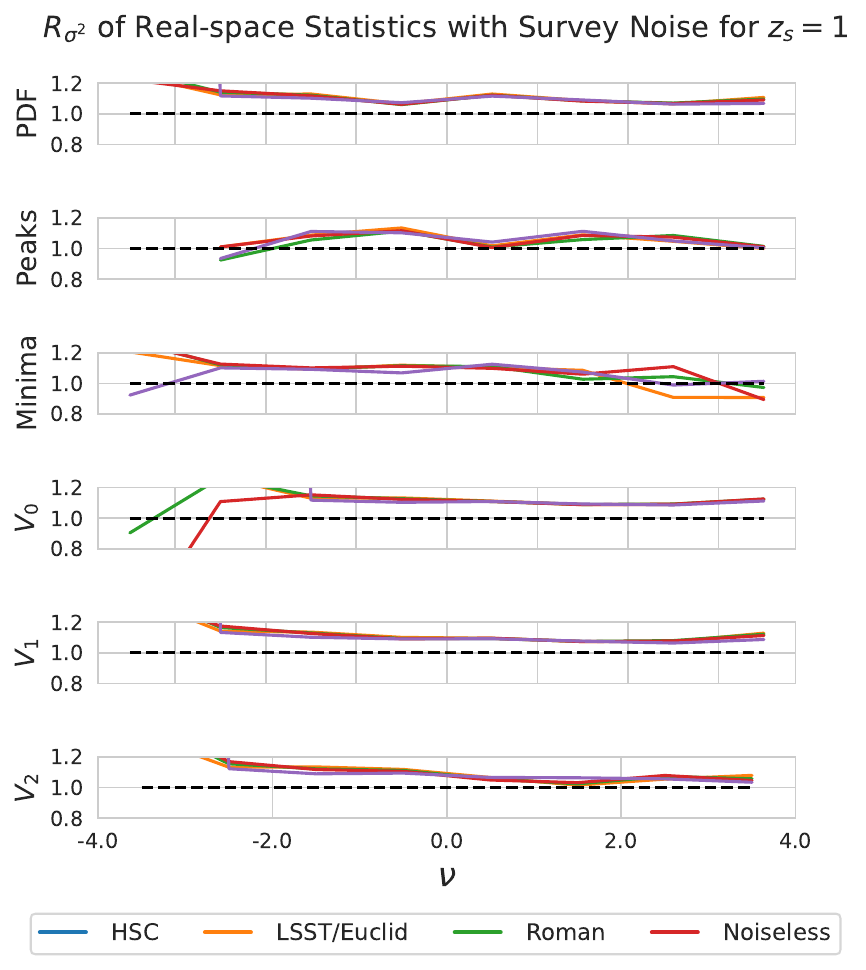} 
    \end{subfigure}    
    \caption
    {
        Ratios of the variance of summary statistics between the BIGBOX and TILED simulations for Replication-Minimal Patches (RMP) at $z_s=1$ for noise levels of various different surveys (given in Table \ref{tab:survey_comparison}). Fourier-based statistics are on the right, and real-space statistics are on the right.
    }
    \label{fig:noise}
\end{figure}

\begin{figure}[t]
    \centering 
    \includegraphics[width=0.6\textwidth]{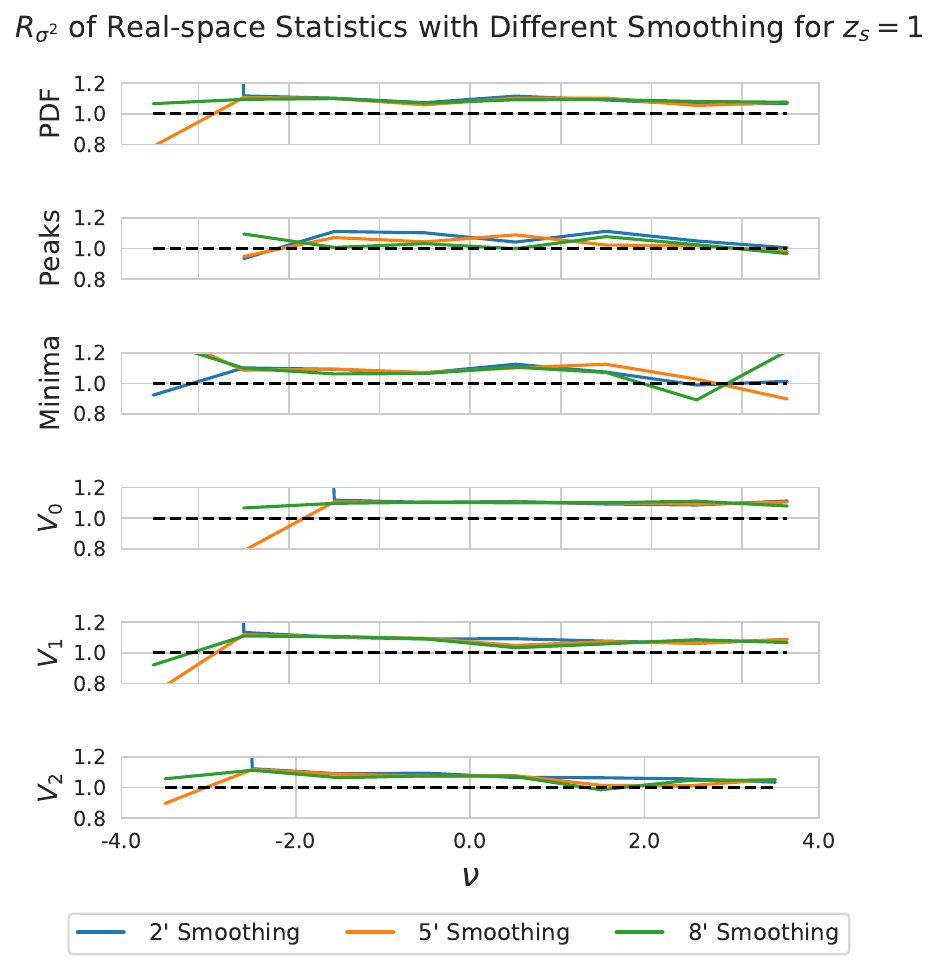} 
    \caption
    {
       Ratios of the variance of summary statistics between the BIGBOX and TILED simulations for Replication-Minimal Patches (RMP) at $z_s=1$ for various different smoothing scales.
    }
    \label{fig:smooth}
\end{figure}

We now study the effects of noise and smoothing scale, having considered noiseless maps with $2'$ smoothing thus far. We specifically focus on RMPs, to investigate the interplay with the super-sample covariance---random noise will impact BIGBOX and TILED boxes in the same way and thus reduce the overall impact of super-sample covariance. We do not show results for DES as they are the most noise dominated and clutter the plot.

Figure \ref{fig:noise} shows the variance ratio for all higher order statistics for the noise levels of various surveys (given in Table \ref{tab:survey_comparison}). For Fourier-based statistics, i.e.~power spectrum and bispectrum, the effects of super-sample covariance is already quite small, thus adding noise does not change the ratio. For $\nu$-bin statistics, the super-sample effects that were present in the noiseless regime still remains after adding noise. Figure \ref{fig:smooth} shows the effects of smoothing, where it can interestingly be seen adding noise does not reduce the super-sample effect at all.

\section{Conclusions}\label{sec:conclusions}
In this work, we investigated the impact of box replication and super-sample effects on the mean and covariance of the power spectrum, bispectrum, PDF, peak/minima counts, and Minkowski functionals.
We utilized weak lensing maps generated from large simulations (``BIGBOX'', $3750\,h^{-1}\mathrm{Mpc}$, produced as part of the HalfDome project \cite{2024arXiv240717462B}) and compared them with those from tiling smaller-box simulations (``TILED'', $625\,h^{-1}\mathrm{Mpc}$) that have the same cosmology and resolution. We study the effects as a function of survey noise level, smoothing, and redshift, for each statistic. 
Our key findings are: 
\begin{itemize} 
    \item Using a Fibonacci splitting of the sphere and gnomonic projection can lead to large $>10\%$ bias in summary statistics due to replication when a patch is centered along a direction of replication. We removed these rare patches from the rest of our analysis.
    \item Box replication artifacts lead to biases of $\mathcal{O}(10\%)$ in the mean of the PDF and Minkowski functionals, and $\mathcal{O}(1\%)$ for all other summary statistics along directions of the sky that particularly contain repeated structures (denoted RIPs). Covariance and correlation ratios in these patches deviate by $\mathcal{O}(10\%)$, indicating spurious replication-induced correlations.
    \item After removing RIPs to better investigate the impact of super-sample effects, there is only a $\leq 1\%$ bias across all statistics, redshifts, and multipole ranges. However, variances differ by 10-30\%, with the largest deviations occurring at high multipoles and redshifts. 
    \item When considering shape noise, ongoing and upcoming surveys such as DES, HSC, KiDS, LSST, Euclid, and Roman are found to also have super-sample covariance effects of 10-25\%, particularly in high redshift bins. That is to say, the shape noise in any of these surveys is not enough to remove the super-sample effect, particularly in $\nu$-binned statistics. Although, splitting into multiple tomographic bins could increase the shape noise to an extent that it dominates the SSC.
    \item Varying the Gaussian smoothing scale also does not remove the super-sample effect, except for tempering the $\nu$-binned statistics at low $\nu$.
\end{itemize}

Our work shows that tiling small box simulations can be a cost-effective way to model the mean of WL statistics, provided that RIP patches are identified and cut out. However, to accurately capture covariances, large box simulations are necessary to avoid both replication and super-sample effects, in particular for small scales and high redshifts.
Our results are for a maximum source redshift $z_s = 2.5$ and a maximum multipole of $\ell=3000$ (for the power spectrum and bispectrum) -- artifacts will increase in magnitude at higher redshifts and smaller scales.

The tiled boxes in our work have size $625\,h^{-1}{\rm Mpc}$---a common choice for weak lensing analyses---but it is instructive to consider how the effects reported in this paper scale with the size of the tiled box as next-generation simulations start using larger boxes (e.g.~$1250\,h^{-1}{\rm Mpc}$ \cite{DES:2024xij}). 
In terms of replication, for weak-lensing light-cones the covariance correction is expected to scale roughly between $L^{-2}$ and $L^{-3}$, depending on the ratio of the light-cone cross-section to the box size, so doubling the box size typically reduces replication-induced covariance by a factor of $\sim 4\text{--}8$ at a given redshift.
In terms of the super-sample effect, the SSC is known to scale as 
\begin{equation} \label{eqn:sigmab}
    \sigma_b^2 \equiv \int \frac{d^3 {k}}{(2\pi)^3} | W({k})|^2 P_{\rm lin}(k),
\end{equation}
where $W(k)$ is the window function (normalized such that $W(0)=1$ and taken as a top hat of length equal to the simulation box size) and $P_{\rm lin}(k)$ is the linear matter power spectrum \cite{PhysRevD.87.123504}. This has also been verified for higher-order statistics \cite{PhysRevD.108.043521}. A box of length $L$, and using $P_{\rm lin}(k)\sim k^n$ for intuition, gives $\sigma_b^2\sim 1/L^{3+n}\sim L^{-4}$ (taking $n\sim1$ which is appropriate for the large scales that dominate the SSC). Computing $\sigma_b$ exactly, using the true linear power spectrum for equation (\ref{eqn:sigmab}), $\sigma_b^2$ is $13.2$ smaller for $1250\,h^{-1}{\rm Mpc}$ than for $625\,h^{-1}{\rm Mpc}$, implying an SSC effect of $\sim20\%$ becomes $\sim1.5\%$.

In this work we stacked boxes periodically---a strategy that preserves boundary continuity but maximises replication artifacts. One way to reduce replication effects is by rotating individual mass shells---as implemented in \cite{2024MNRAS.530.5030O}---this suppresses direct repetition but introduces sharp discontinuities at the boundaries that can contaminate higher-order statistics in other ways. 
A systematic comparison of these methods, and of hybrid schemes that trade replication artifacts for discontinuities, is an important direction for future study, but in either case, super-sample effects will remain. 
Finally, while not relevant for full sky simulations, another common approach to simulating weak lensing light-cone patches is to use a hierarchy of slices from boxes with different sizes along the light-cone \cite{Sato_2009}: often using box sizes less than $625\,h^{-1}{\rm Mpc}$ towards the low redshift end of the light-cone and larger boxes at high redshifts. This setup is also missing super-sample modes---different super-sample modes to the setup considered in our work---and thus could lead to a different super-sample effect; nevertheless, this setup was  considered analytically for the $C_\ell$ SSC in \cite{2018JCAP...06..015B}, finding a similar order of magnitude effect as found in our work. A comparison of the super-sample effect for higher-order statistics using this setup would be interesting future work.

With large efforts being made to produce full sky simulations \cite{2024MNRAS.530.5030O, 2024arXiv240717462B} for weak lensing, CMB lensing, and other observables in the sky, it is also important to consider the implications of our results for a full-sky analysis. Repeating the analysis of this paper by directly computing covariances for a full sky simulation would require thousands of full-sky simulations, both tiled and not-tiled, which currently are not available. Nonetheless, our results provide motivation for the need of large box simulations for accurate full-sky analyses. While the projection effects we found will not be relevant, provided one analyses the full sky while respecting its spherical geometry, the replication effects we observed in certain patches of the sky will be present throughout the full sky, and super-sample effects will also be present. Moreover, \cite{2023OJAp....6E...1U} showed that considering $\kappa-\mu_\kappa$ patches gives the same SSC behavior as the full sky, which we show in Appendix \ref{app:nu_kap} is also a $\mathcal{O}(10\%)$ effect.
The increase in replication and super-sample effects with source redshift implies particular importance for CMB lensing, which has $z_s=1{,}100$.
Moreover, while we considered weak lensing in this paper, it would be fruitful future work to use correlated simulations \cite{2024MNRAS.530.5030O, 2024arXiv240717462B} to study the impact of replication and super-sample effects on the combined analysis of different observables, such as galaxy clustering, CMB secondaries, and fast radio bursts (see e.g.~\cite{Konietzka:2025kdr} for a discussion of replication effects in the context of fast radio bursts).

All lensing maps produced in this study are publicly available as part of the HalfDome project\footnote{\url{https://halfdomesims.github.io/}}.

\section*{Acknowledgements}
We thank Yu Feng, Joachim Harnois-Déraps, Biwei Dai, Joe DeRose, Niall Jeffrey, Zack Li, Uroš Seljak, and Cora Uhlemann for useful discussions.
The HalfDome project is supported by JSPS KAKENHI Grants 
23K13095, 23H00107, 25H00403 (JL). JA is supported
by JSPS KAKENHI Grant JP23K19064. RT is supported by JSPS KAKENHI Grant 23KJ0747. LT was supported by JSPS KAKENHI Grant 24K22878. 
This research used resources of the National Energy Research Scientific Computing Center (NERSC), a U.S. Department of Energy Office of Science User Facility located at Lawrence Berkeley National Laboratory, operated under Contract No. DE-AC02-05CH11231 using NERSC award HEP-ERCAP0023125. 
We acknowledge the Texas Advanced Computing Center (TACC) at The University of Texas at Austin for providing grid resources that have contributed to our work. This work was supported by MEXT as Program for Promoting Researches on the Supercomputer ``Fugaku'' (Multi-wavelength Cosmological Simulations for Next-generation Surveys, JPMXP1020230407) and used computational resources of supercomputer Fugaku provided by the RIKEN Center for Computational Science (Project ID: hp230202).

\appendix
\section*{Appendices}

\begin{figure}[t]
    \centering 
    \includegraphics[width=\textwidth]{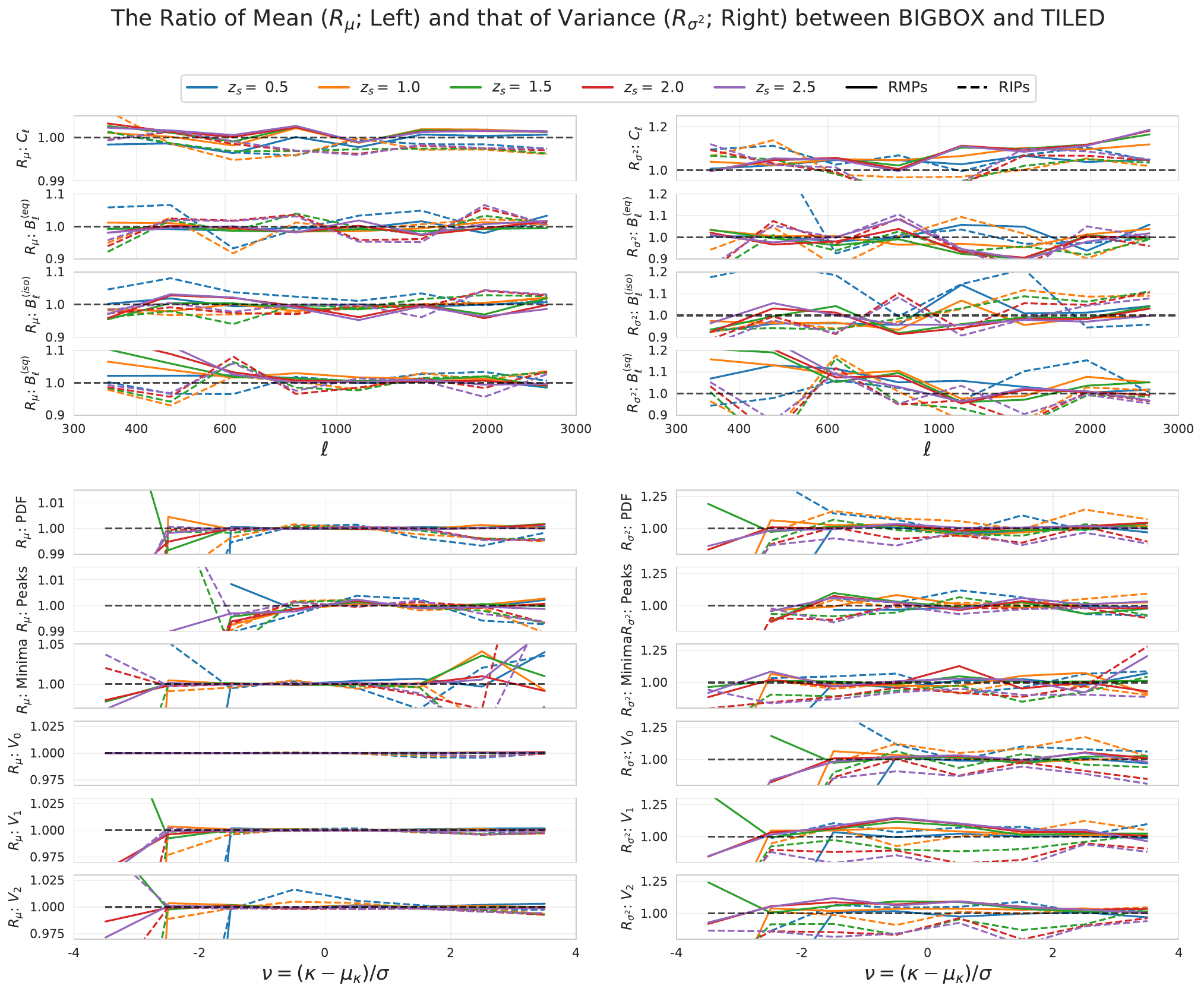} 
    \caption[BIGBOX/TILED Ratios of Mean and Variance for RIPs and RMPs]
    {
        Same as Fig.~\ref{fig:boxreplication_main} but for $\nu\equiv(\kappa-\mu_\kappa)/\sigma$.
    }
    \label{fig:boxreplication_numean}
\end{figure}

\begin{figure}[t]
    \centering 
    \includegraphics[width=\textwidth]{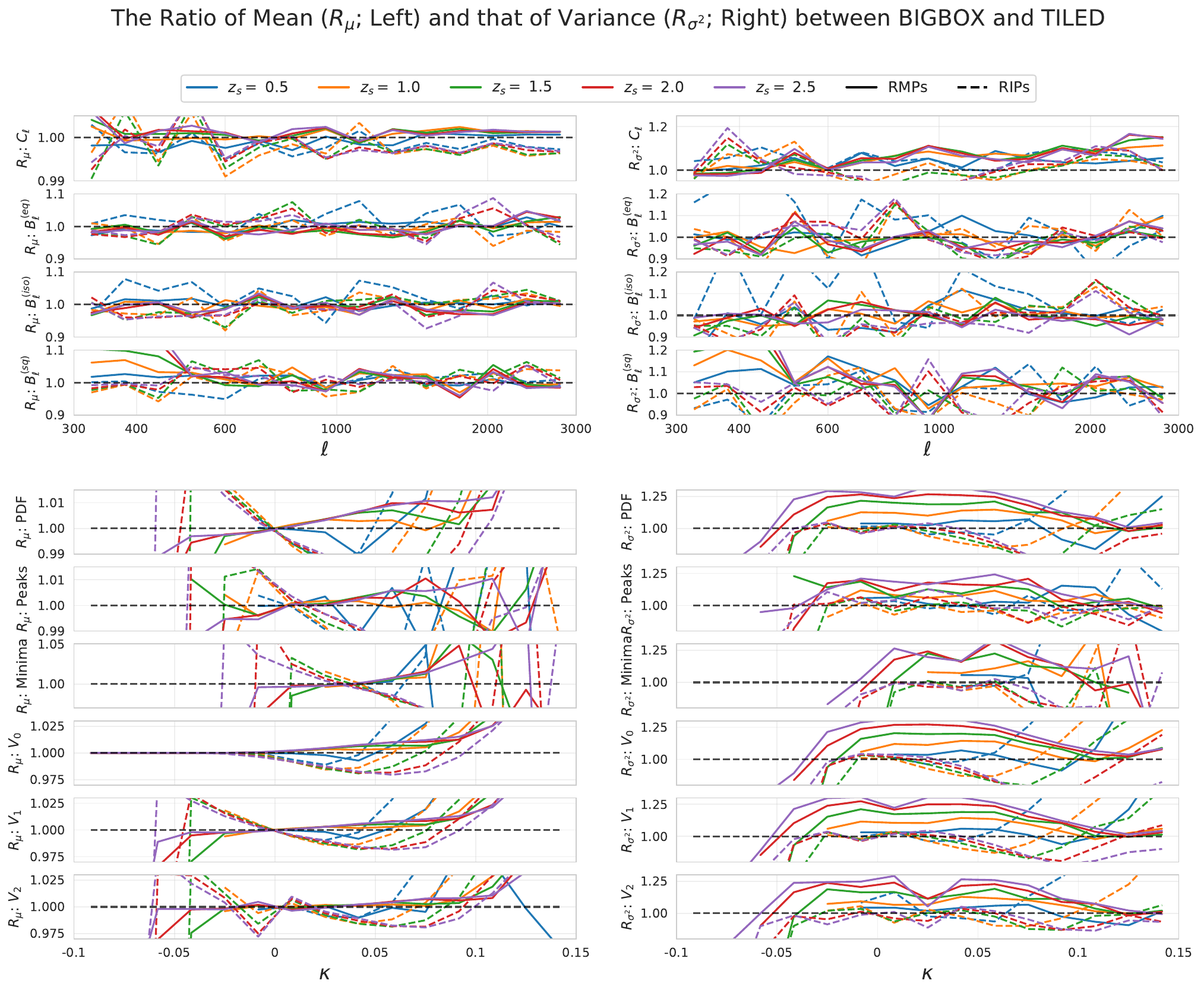} 
    \caption[BIGBOX/TILED Ratios of Mean and Variance for RIPs and RMPs]
    {
        Same as Fig.~\ref{fig:boxreplication_main} but for $\kappa$.
    }
    \label{fig:boxreplication_kap}
\end{figure}

\begin{figure}[t]
    \centering 
    \includegraphics[width=\textwidth]{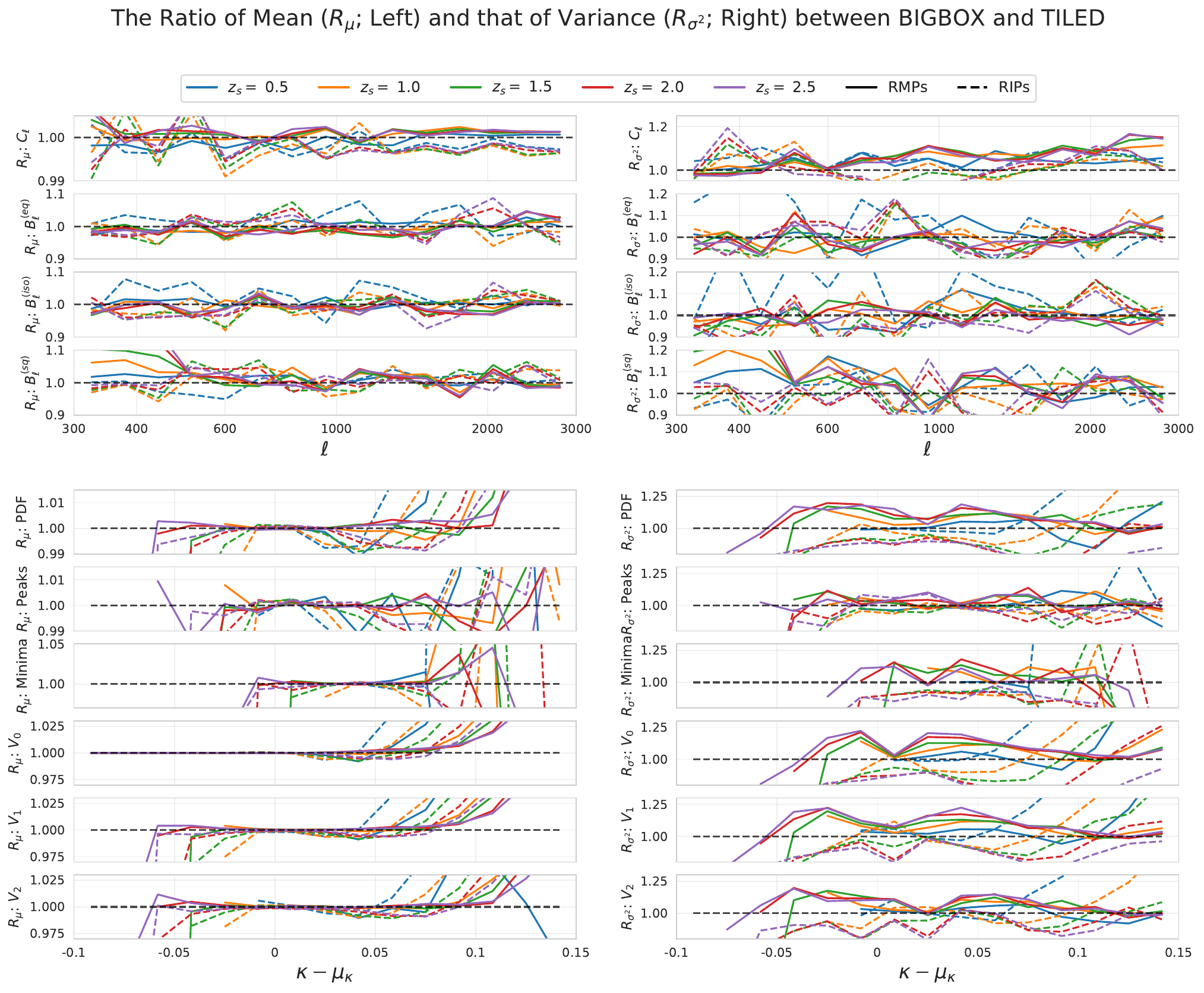} 
    \caption[BIGBOX/TILED Ratios of Mean and Variance for RIPs and RMPs]
    {
        Same as Fig.~\ref{fig:boxreplication_main} but for $\kappa-\mu_\kappa$.
    }
    \label{fig:boxreplication_kapmean}
\end{figure}

\section{Different normalizations the maps} \label{app:nu_kap}

One can normalize $\kappa$ maps in various ways. In the main text (Fig.~\ref{fig:boxreplication_main}) we considered variance scaled maps $\nu\equiv\kappa/\sigma$. We now consider three other combinations: $\nu\equiv(\kappa-\mu_\kappa)/\sigma$, $\kappa$, and $\kappa-\mu_\kappa$ in Figures \ref{fig:boxreplication_numean}, \ref{fig:boxreplication_kap}, and \ref{fig:boxreplication_kapmean} respectively. Whether one normalizes by the variance or subtracts out the local mean naturally impacts the differences between BIGBOX and TILED.

In general, the $\mathcal{O}(10\%)$ effect reported in the main paper remains for all normalization choices, although the effect shifts between scales and patch types in various ways. 
Working in terms of $\kappa$ instead of $\nu$ has a similar magnitude of SSC; moreover, the mean and variance corrections in RIPs are larger when using $\kappa$ instead of $\nu$, and effect a significantly larger range of bins.
Subtracting the mean can be seen to remove cosmic variance induced bias on large scales for the Fourier-based statistics, and also reduce the SSC, however it increases the variance caused by replication. 
It was shown in \cite{2023OJAp....6E...1U} that considering $\kappa-\mu_\kappa$ patches of the PDF gives the same covariance behavior as the full-sky PDF, implying that the $\mathcal{O}(10\%)$ effects are relevant for the full sky.

\section{Covariance and Correlation Matrices} \label{app:cor}

\subsection{Replication Effect} \label{app:cor_rep}

\begin{figure}[p]
    \centering
    \includegraphics[width=\textwidth]{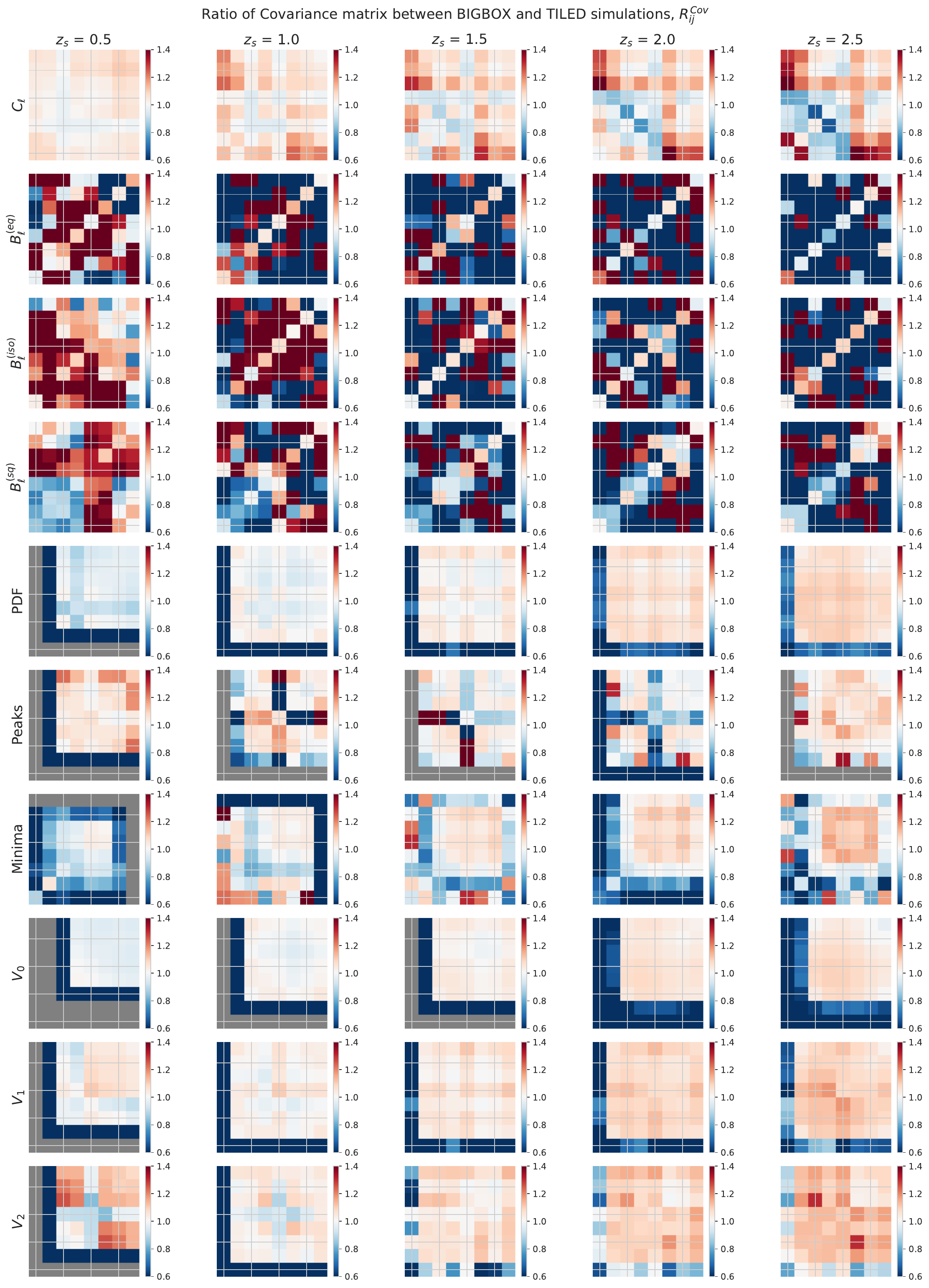}
    \caption[Covariance Ratios for RIP]
    {
        BIGBOX/TILED ratios of covariance matrices specifically for RIPs. Covariance structures unique to RIPs, such as \(C_\ell\) and \(V_0\), indicate that box replication artifacts can generate distinct covariance patterns.
    }
    \label{fig:boxreplication_cov_RIP}
\end{figure}

\begin{figure}[p]
    \centering
    \includegraphics[width=\textwidth]{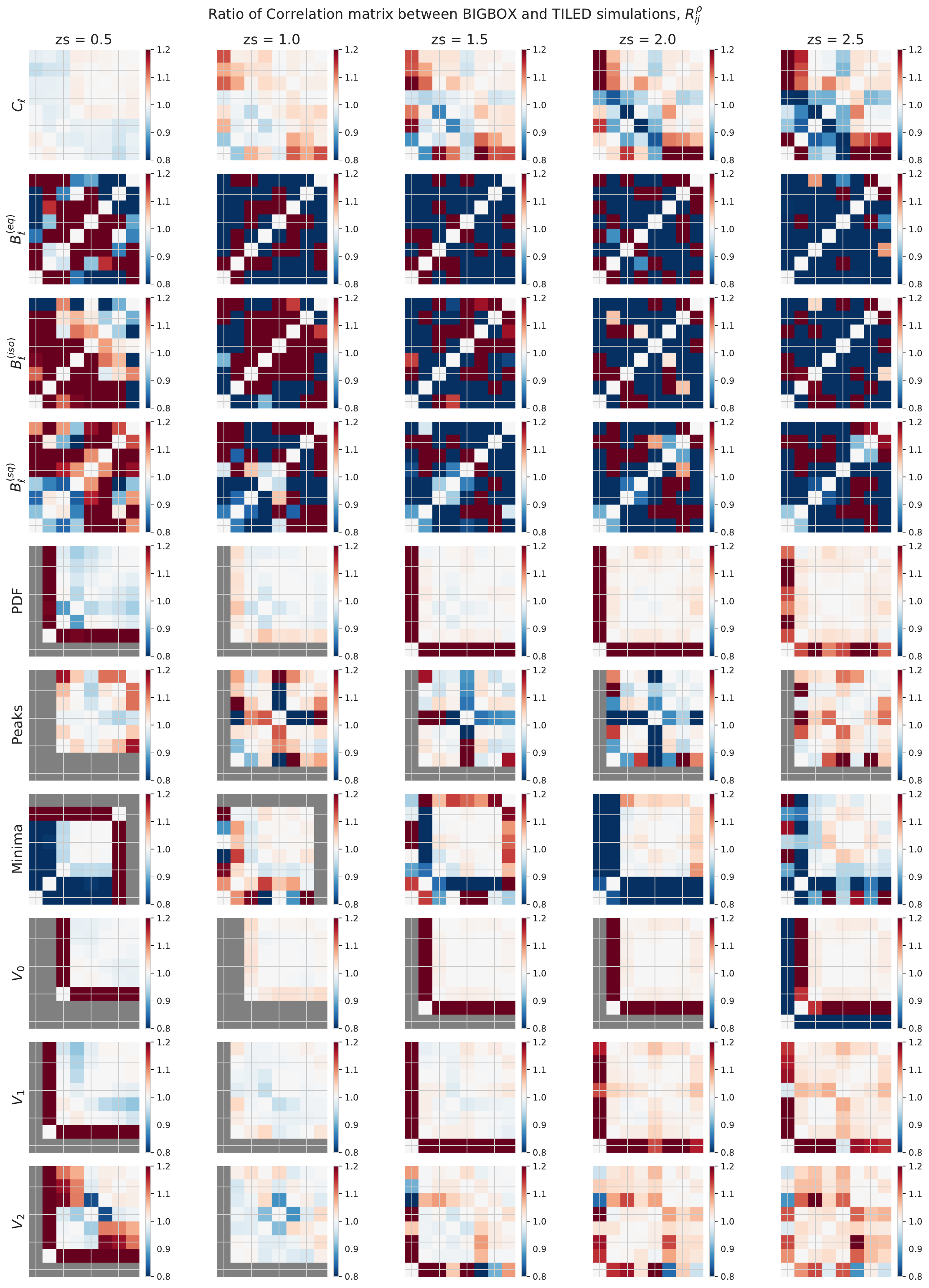}
    \caption[Correlation Ratios for RIP]
    {
        BIGBOX/TILED ratios of correlation matrices specifically for RIPs. Similar to the covariance ratios, RIP correlation ratios are consistently lower than those for RMPs. Off-diagonal structures tend to converge toward unity at higher redshifts.
    }
    \label{fig:boxreplication_corr_RIP}
\end{figure}

Figures~\ref{fig:boxreplication_cov_RIP} and~\ref{fig:boxreplication_corr_RIP} display the ratios of covariance and correlation matrices between BIGBOX and TILED simulations for RIPs. Our main observations are that RIPs consistently exhibit lower covariance ratios compared to RMPs, and that the correlation ratios for RIPs are generally lower than those for RMPs—except at \(z_s=0.5\), where the super-sample effect is present in both TILED and BIGBOX simulations. Furthermore, distinct covariance structures, such as those for \(C_\ell\) and \(V_0\), emerge due to box replication artifacts. Structural transitions observed in previous analyses occur at higher redshifts (\(z_s \approx 2.0\sim2.5\)) in RIPs, aligning with RMPs as higher redshift regions become dominated by RMPs. 

\subsection{Super-Sample Effect} \label{app:cor_ssc}

\begin{figure}[p]
    \centering
    \includegraphics[width=0.98\textwidth]{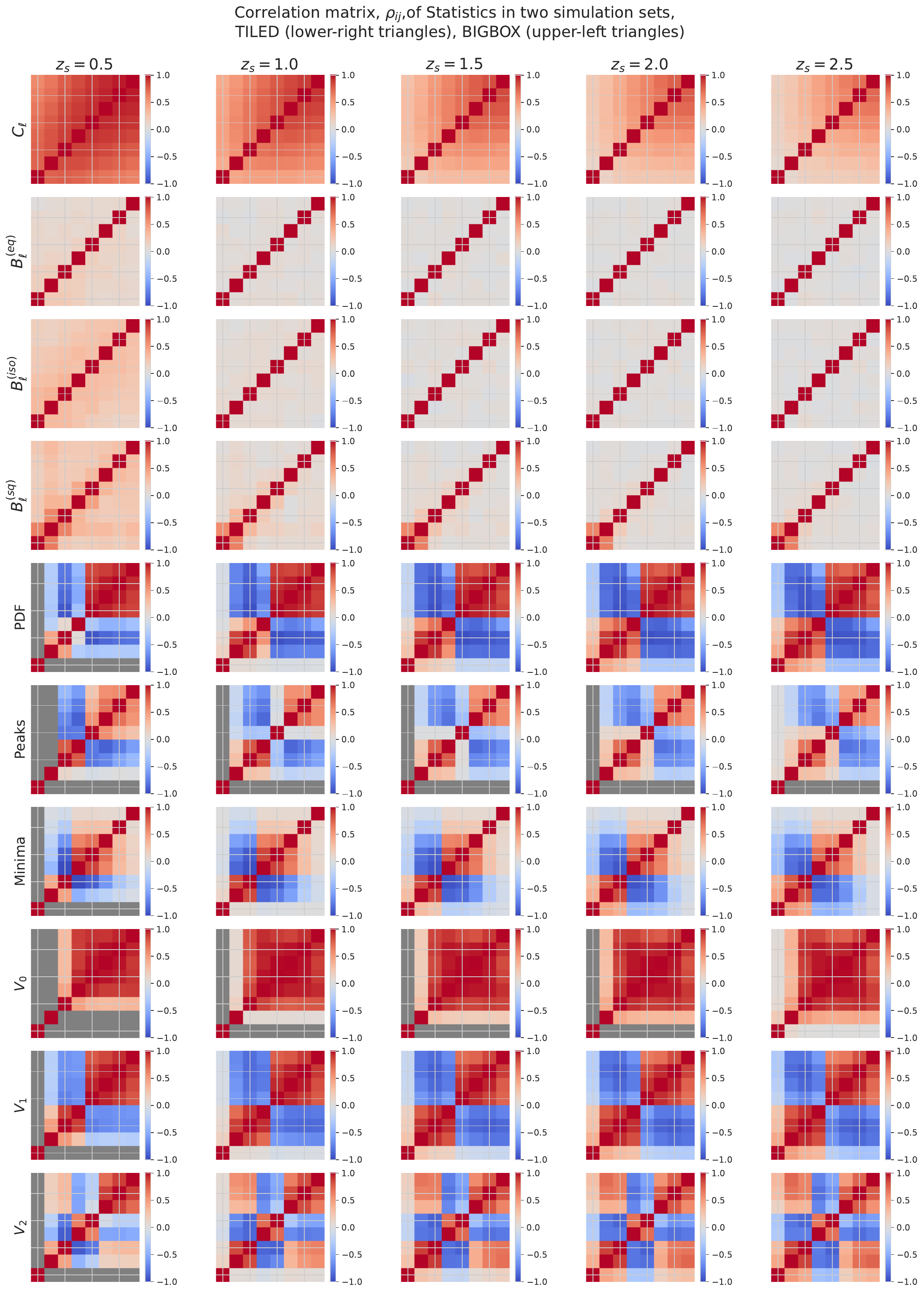}
    \caption[Correlation Matrices of Summary Statistics in BIGBOX and TILED Simulations]
    {
        Correlation matrices of various summary statistics in the BIGBOX and TILED simulations across multiple source redshifts ($z_s = 0.5$, $1.0$, $1.5$, $2.0$, and $2.5$). The upper-left triangle of each panel displays the correlation coefficients from the BIGBOX simulations, while the lower-right triangle shows the corresponding values from the TILED simulations. The similar color scales indicate that the correlation structures are almost consistent between the two simulation sets.
    }
    \label{fig:corr_main}
\end{figure}

\begin{figure}[p]
    \centering
    \includegraphics[width=0.98\textwidth]{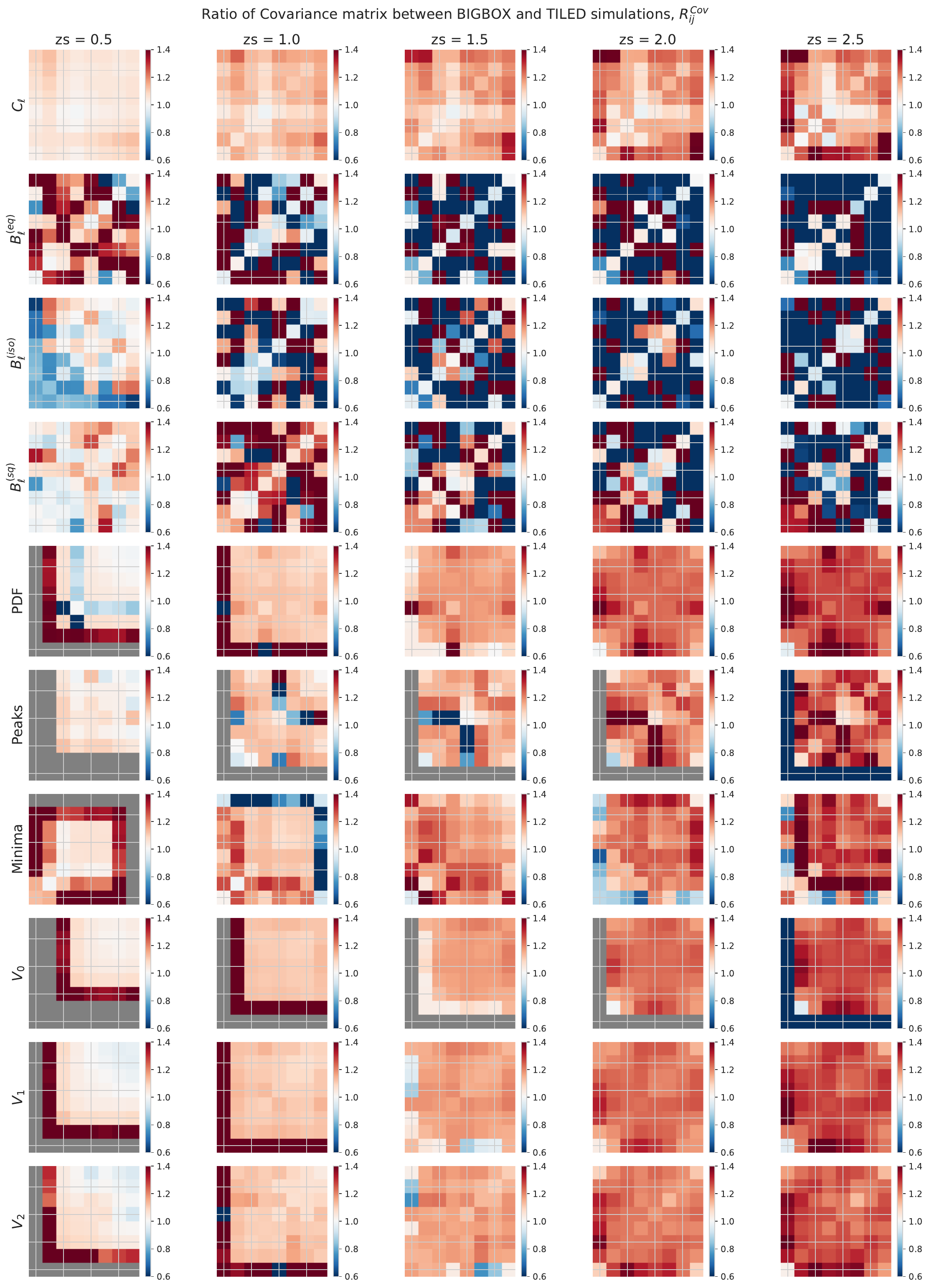}
    \caption[Ratios of Covariance Matrices between BIGBOX and TILED Simulations]
    {
        Ratios of covariance matrices ($R_{\text{Cov}}$) between the BIGBOX and TILED simulations across multiple source redshifts ($z_s = 0.5$, $1.0$, $1.5$, $2.0$, and $2.5$). The ratios consistently range from $10\%$ to $30\%$ above unity for most measured statistics, indicating the substantial impact of super-sample effect on the covariance structure.
    }
    \label{fig:cov_ratio_main}
\end{figure}

\begin{figure}[t]
    \centering
    \includegraphics[width=0.98\textwidth]{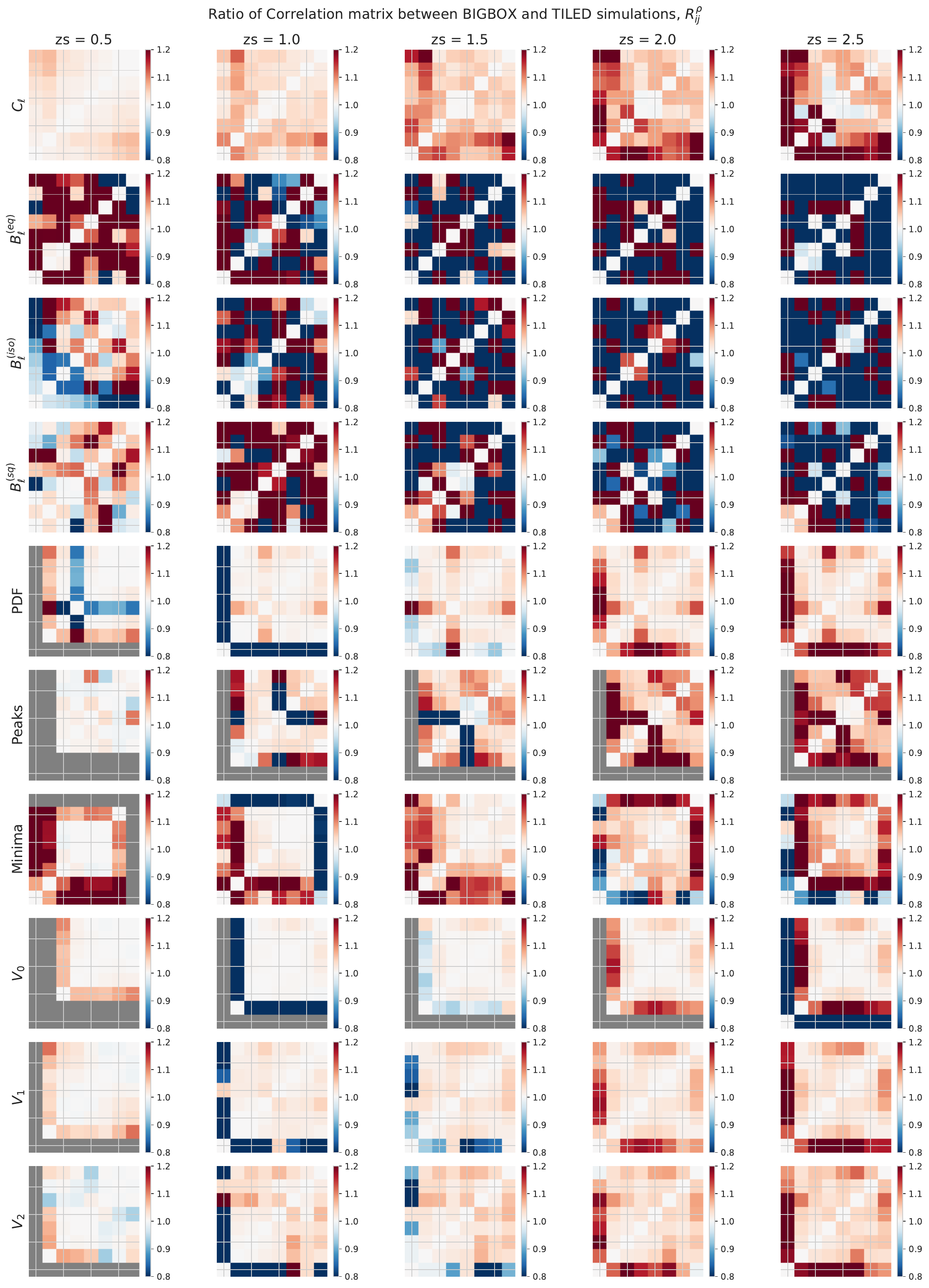}
    \caption[Ratios of Correlation Matrices between BIGBOX and TILED Simulations]
    {
        Ratios of correlation matrices ($R_{\rho}$) between the BIGBOX and TILED simulations across multiple source redshifts ($z_s = 0.5$, $1.0$, $1.5$, $2.0$, and $2.5$). The ratios generally deviate by less than $5\%$, indicating that while the super-sample effect elevates the entire covariance structure, its impact on the correlation coefficients remains minimal.
    }
    \label{fig:corr_ratio_main}
\end{figure}

In this subsection, we quantify the impact of the super-sample effect on the covariance and correlation matrices of various summary statistics derived from the BIGBOX and TILED simulations. Figure~\ref{fig:corr_main} shows the corresponding correlation matrices, with the upper-left triangle representing the BIGBOX results and the lower-right triangle displaying the TILED results. The similar color scale indicates that the correlation structures are nearly identical between the two simulation sets. Figures~\ref{fig:cov_ratio_main} and~\ref{fig:corr_ratio_main} present the ratios of the covariance and correlation matrices—denoted as $R_{\text{Cov}}$ and $R_{\rho}$, respectively—which provide a quantitative measure of the super-sample effect's influence on both the diagonal and off-diagonal elements of the covariance structure. 

The covariance ratios $R_{\text{Cov}}$ generally exceed unity, ranging from 10\% to 30\% above the baseline values for most summary statistics, underscoring the significant contribution of the super-sample effect to the overall covariance matrix. Conversely, the correlation ratios $R_{\rho}$ typically deviate by less than 5\%, indicating that while the super-sample effect elevates the covariance, its impact on the correlation coefficients is smaller. Notable exceptions are observed in the angular power spectrum and in the peak and minima counts. For the angular power spectrum, larger deviations in $R_{\rho}$ are consistent with the super-sample effect, which introduces strong correlations through shared large-scale modes, particularly at small multipole moments ($\ell$). In the case of peak and minima counts, the deviations in correlation ratios likely result from systematic shifts in peak positions in the TILED simulations, effecting correlations near the peak bins.

\bibliographystyle{JHEP}
\bibliography{main.bib} 

\end{document}